\documentclass{article}
\usepackage[utf8]{inputenc}
\usepackage{amssymb}
\usepackage{caption}
\usepackage{algorithm}
\usepackage{algpseudocode}
\algrenewcommand\alglinenumber[1]{\scriptsize #1:}
\usepackage{graphicx}

\title{Random Expansion Method for the Generation of Complex Cellular Automata}
\author{Juan Carlos Seck-Tuoh-Mora*, Norberto Hernandez-Romero, \\
Joselito Medina-Marin, Genaro J. Martinez, \\
Irving Barragan-Vite\\
AAI-ICBI-UAEH. Carr Pachuca-Tulancingo Km 4.5. \\
Pachuca 42184 Hidalgo. Mexico\\
Unconventional Computing Centre, University of \\
the West of England, BS16 1QY Bristol, United Kingdom\\
Escuela Superior de Computo, Instituto Politecnico \\
Nacional, Mexico\\}
\date{September 2020}

\begin{document}

\maketitle

\begin{abstract}
The emergence of complex behaviors in cellular automata is an area that has been widely developed in recent years with the intention to generate and analyze automata that produce space-moving patterns or gliders that interact in a periodic background. Frequently, this type of automata has been found through either an exhaustive search or a meticulous construction of the evolution rule. In this study, the specification of cellular automata with complex behaviors was obtained by utilizing randomly generated specimens. In particular, it proposed that a cellular automaton of $n$ states should be specified at random and then extended to another automaton with a higher number of states so that the original automaton operates as a periodic background where the additional states serve to define the gliders. Moreover, this study presented an explanation of this method. Furthermore, the random way of defining complex cellular automata was studied by using mean-field approximations for various states and local entropy measures. This specification was refined with a genetic algorithm to obtain specimens with a higher degree of complexity. With this methodology, it was possible to generate complex automata with hundreds of states, demonstrating that randomly defined local interactions with multiple states can construct complexity.
\end{abstract}

Keywords: Cellular automata, complexity, mean-field theory, local information, entropy

Submitted to: Inf Sci

\section{Introduction}
\label{intro}

Given that the emergence of complexity occurs through simple local interactions between the automaton cells, complexity has been one of the most investigated topics in the study of cellular automata \cite{mitchell2009}. Researchers have intended to study automata that are capable of creating mobile structures (known as gliders) that interact in a periodic background. These structures do not disintegrate or end up dominating the space of evolutions but rather maintain a balance with the periodic background while interacting with each other \cite{bar2019}. This is the type of cellular automata that has been treated in this study. In particular, the one-dimensional case has been analyzed here.

The study of gliders in cellular automata has been a popular field of research since an enormous amount of work began on the study of Conway's Game-of-Life cellular automaton. Some excellent studies on this automaton were the books edited by Griffeath and Moore \cite{griffeath2003new} and Adamatzky \cite{adamatzky2010game}. For the one-dimensional case, a boom was caused by Wolfram's study \cite{wolfram2002new} \cite{wolfram2018cellular}. The implications of the generation of gliders in cellular automata peaked with the study conducted by Cook, who, through the interaction of gliders, demonstrated that the cellular automaton Rule 110 is universal \cite{cook2004universality}. 

Multiple ways of generating complex cellular automata have been proposed and developed. We classify these methods in four main categories. However, these categories are not exclusive, and a study can simultaneously fall into more than one of these at once. Given the impossibility of referring to all the works related to the generation of complex cellular automata, we presented only a subset of the most relevant and recent studies in this field of research. Consequently, many other significant works have been omitted here.

The first category is represented by the studies that used a more theoretical approach regarding the application of mathematical concepts. In this category, we include studies such as  \cite{ISI:000447313200004}, which presented a cellular automaton whose evolution rule was an approximation of two-dimensional Kolmogorov complexity. The automaton was shown to be capable of simulating binary logic circuits. In the same study, a similar cellular automaton with increasing complexity produced gliders that could be used as information carriers. Besides, a glider gun and logic gates were constructed as well. A probabilistic analysis that enabled us to predict the global behavior of configuration patterns in elementary cellular automata was presented in \cite{ISI:000236886100043}, for which both the states were quiescent, from asynchronism to full synchronism. With this, five automata exhibited complex phase transitions. The phase transitions in asynchronous cellular automata were investigated in \cite{ ISI:000365183600002} by using the local structure theory in order to estimate different types of second-order phase transitions. In \cite{jin2014glider} and \cite{jin2016symbolic}, a mathematical definition of gliders for one-dimensional cellular automata was presented to provide a symbolic dynamics characterization demonstrating that glider interaction implies chaos in the sense of Li-Yorke. Moreover, ultradiscretization was applied in \cite{murata2015multidimensional} in order to transform the multidimensional Allen-Cahn equation into a cellular automaton, obtaining traveling wave solutions that were similar to those in the continuous systems. In \cite{bidlo2016routine}, logical and algebraic properties were used to construct conditionally matching evolution rules in one and two dimensions to design replicating loops and square calculation tasks. In \cite{ISI:000386576900001}, the minimal Boolean form of binary cellular automata was applied to characterize their complexity. New tools to study self-organization in cellular automata with gliders were introduced in \cite{de2017self}. Initial configurations were defined according to an ergodic measure, and the limit measure described the asymptotic behaviors of gliders. Mean-field theory, de Bruijn and subset diagrams, and memory were used in \cite{martinez2019patterns} in order to display nontrivial constructions and quasi-chaotic behavior in elementary cellular automaton Rule 22.

Another category comprises of studies that proposed evolution rules inspired by the physical, chemical, and biological phenomena. For instance, unconventional computer systems inspired by  collision-based computing in cellular automata that work through the interaction of gliders in a periodic background was investigated in \cite{adamatzky2012collision}. Moreover, the ultradiscretization of reaction-diffusion partial differential equations was developed in \cite{ohmori2015cellular} in order to obtain evolution rules with dynamical properties such as bistability, pulse annihilation, soliton-like preservation, and periodic pulse generation. A three-state hexagonal cellular automaton was studied in \cite{adamatzky2006glider}, which was a discrete model of a reaction-diffusion system with inhibitor and activator reagents. The cellular automaton exhibited gliders used to implement basic computational operations, and, thus, collision-based logical universality was demonstrated. Moreover, Actin cellular automata were presented in \cite{adamatzky2015actin}, which were based on the behavior of a globular protein that forms long filaments for intracellular signaling. The model consisted of two binary-state semi-totalistic automaton arrays that support gliders. Some properties were predicted using Shannon entropy. In \cite{alonso2016actin}, actin cellular automata were enriched with memory. This additional feature slowed the propagation of gliders down, decreasing entropy and transforming some gliders to stationary oscillators and stationary oscillations to still patterns. Furthermore, in \cite{dourvas2017cellular}, the simplicity of cellular automata was combined with a light-sensitive version of Belousov-Zhabotinsky reaction. The resulting model could simulate a one-bit full adder digital component. Three natural processes (spatial competition and the distinction between the stronger and weaker agents as well as inertia) were studied in \cite{kramer2017emergence} in order to propose a cellular automaton model in which spatial patterns resembled phase transitions, jumps, and discontinuous transitions. Additionally, the morphological complexity of the Gray-Scott reaction-diffusion system was studied in \cite{adamatzky2018generative} through the use of Shannon entropy, Simpson diversity, Lempel-Ziv complexity, and expressivity to exhibit that the Gray-Scott systems support wave-fragments and gliders. In \cite{dourvas2018inhibitor}, a cellular automaton with wave propagation characteristics on an excitable medium was presented. The automaton also used inhibitors to create a simplified basic transistor. A combination of two of those transistors could reproduce universal logic gates. Besides, a lattice-gas cellular automaton was presented in \cite{fates2018trade} in order to mimic the self-organization process of swarm formation. According to the parameters that define the automaton, the system may display a great variety of patterns. Excitable cellular automata were studied in \cite{mayne2019cellular} in order to simulate signal transmission in actin networks sampled from slime mold Physarum polycephalum. Actin networks support directional transmission to implement Boolean logical operations for designing actin circuitry. Actin cellular automata were investigated in \cite{adamatzky2019discovering}, where two-state transition rules were considered: a Game-of-Life-like automaton with two states and an excitable actin automaton with three states. Eight-argument Boolean functions were also used to implement the logical functions.

A third category is represented by studies that analyzed the modifications of the traditional cellular automaton model. For instance, elementary cellular automata with memory capabilities were presented in \cite{alonso2005elementary}, where several patterns showing glider interaction were obtained. In \cite{martinez2012complex}, the majority memory can generate gliders in elementary cellular automata with chaotic behavior. In \cite{martinez2012soliton}, the property of solitonic collisions of gliders was studied in elementary cellular automata. Moreover, memory was also employed in \cite{martinez2010make} and \cite{martinez2015dynamics} to create complex behaviors and provide a classification of the dynamics in Rule 126 with memory. Hybridization is another modification investigated in these studies. Specifically, a wide range of gliders was studied in \cite{chen2015glider} and \cite{chen2016glider}, taking hybrid cellular automata whose evolution was dependent on two or more evolution rules. For example, Rules 168 and 133 produced complicated interactions that were classified by using a quantitative approach. In \cite{lee2016characterization}, a computationally universal Brownian cellular automaton was described based on an asynchronous cellular automaton with local cell state transitions following a Poisson point process, which demonstrates that random fluctuations can serve for efficient computation. The dynamics of randomized local neighborhood connections in two-dimensional cellular automata were investigated in \cite{wuensche2018pulsing}. With the three states, the automata held sustained rhythmic oscillations and glider dynamics. Using a totalistic cellular automaton, Game-of-Life-type rules were investigated in \cite{ishida2018possibility}, in which complexity emerged from the interactions between an activating factor and an inhibiting factor. Signed majority cellular automata were investigated in \cite{goles2018complexity} to simulate different types of logic circuitry. It has been proven that uniform asymmetric and non-uniform symmetric rules are universal.
In \cite{morita2019universal}, a simple triangular partitioned cellular automaton was proposed, where the next state of a cell was determined by the three adjacent parts of its neighbor cells. The evolution of some specimens with these features demonstrated the existence of gliders and glider guns. Memory was used in \cite{martinez2018conservative} to simulate Fredkin gates in a one-dimensional cellular automaton by the collision of gliders, solitons, and binary interactions to obtain the final outputs. An algorithm was proposed in \cite{martinez2019universal} to convert any Turing machine into a one-dimensional cellular automaton with a 2-linear time dynamics. Moreover, three Turing machines were converted into three cellular automata: binary sum, Rule 110 and a universal reversible Turing machine.

A fourth category consisted of studies involved in systematic searches using different types of heuristic approaches in order to find complex cellular automata. For instance, in \cite{ISI:000386576900001}, behavioral metrics were employed in a genetic algorithm to select cellular automata similar to the Game of Life, while the spontaneous emergence of glider guns in cellular automata with two states in two dimensions was developed in \cite{sapin2009genetic} and \cite{sapin2010stochastic}. Those works applied an evolutionary search for new glider guns, and an automatic process was provided to classify glider guns that could implement collision-based universal cellular automata. A surface and histogram-based classification for the periodic, chaotic and complex elementary cellular automata was presented in \cite{seck2014emergence} by using the nearest-neighbor interpolation in order to analyze a diversity of surfaces. 

Previous studies demonstrated that the specification of complex cellular automata has been proposed by a meticulous construction of evolution rules, modification of the classical model or an exhaustive search of the suitable specimens. However, thus far, there has been no method to generate cellular automata with gliders that is general for any number of states and can consider the interaction of many states. 

This study presents an original method for the generation of cellular automata in one dimension with hundreds of states by using randomly generated automata as its basis. These automata serve as support for generating a periodic background on which the random extensions on the said automata are capable of generating gliders. With this process, it is possible to obtain systematically complex cellular automata with hundreds of states. The contribution of this study lies in demonstrating that simple local interactions, that are specified at random and involve a large number of states, can achieve complexity (glider generation).

\section{Basic Concepts of Cellular Automata}
\label{basics}

A cellular automaton $\mathcal{A} = \{K, \phi, v \} $ consists of a finite set of states $ K $, a vector neighborhood of relative positions $ v = \{v_1, v_2 \} $, with $ v_i \in \mathbb{Z} $, $ v_1 \leq v_2 $, and an evolution rule $ \phi: K^{(v_2-v_1 + 1)} \rightarrow K $ that maps blocks of states of size $ v_2-v_1 + 1$ to individual states of $ K $. The dynamics of $ \mathcal{A} $ is defined by taking an initial configuration of $ m $ states $ c^0: \mathbb{Z}^m \rightarrow K $, where $ \mathbb{Z}^m $ is the set of integers from $ 0 $ to $m-1$. For $ 0 \leq i \leq m-1 $, a neighborhood $ n_i^0 = c_ {i + v_1}^0 \ldots c_{i + v_2}^0 $ is generated by taking periodic boundary conditions. This is how $ c_i^1 = \phi(n_i ^ 0) $ is defined. In general, $ c_i^j = \phi(n_i^{j-1}) = \phi(c_{i + v_1}^{j-1} \ldots c_{i + v_2}^{j-1})$.

Thus, evolution rule $ \phi $ induces a global mapping $ \Phi: c^j \rightarrow c^{j + 1}$ based on local interactions between the states of the cells of $ c^j $ . 

In order to simplify the study, only cellular automata with neighborhood size $ 2 $ were considered, i.e., $ v = \{v_1, v_1 + 1 \} $, since all the other cases can be simulated to this representation by using more states, as explained below.

Since the evolution rule maps a neighborhood of size $v_2-v_1 + 1 $ to an individual state, the extension of this mapping to $ v_2-v_1 $ neighborhoods will produce a block of $ v_2-v_1 $ states, where the extended vector of neighbor cells is specified as $ v'= \{v_1, 2v_2-v_1-1 \} $. This vector contemplates $ 2 (v_2-v_1) $ states; therefore, this extended mapping can be represented as $ K^{2 (v_2-v_1)} \rightarrow K^{v_2-v_1} $. This means that a mapping of two blocks of size $ v_2-v_1 $ evolves into a block of the same size.

Thus, we can define a new set of states $ S $ such that $ | S | = K^{{2 (v_2-v_1)}} $, along with defining a new evolution rule $ \varphi: S^2 \rightarrow S $ such that $ \varphi $ simulates the same behavior as $\phi$ in the original automaton.

This demonstrates that every cellular automaton in one dimension can be simulated by another $ \mathcal {A} = \{S, \varphi, v \}$ with $v = \{-1,0 \}$, which also entails an exponential increase in the number of states to perform the simulation.

This way, we only consider cellular automata with a neighborhood size $2$, since the other cases can be reduced to this. For this type of automaton, the evolution rule $ \varphi $ can be represented by a matrix $ M_\varphi $ of order $ | S | \times | S | $ such that for any two states $ s,y \in S $,  $ M_\varphi(s, y) = \varphi (sy) $, i.e., each entry of $ M_\varphi $ is determined by the evolution of the specified neighborhood by concatenating its respective row and column indices.

\section{Local Entropy Measures}
\label{entropias}

Several studies have explored various ways of generating and measuring the production of gliders in cellular automata by analyzing the densities \cite{bagnoli2018phase},mean-field theory \cite{martinez2019patterns} and entropy \cite{marr2005topology}, along with reviewing the emergence of structures in the evolution space \cite{hordijk2001upper} \cite{martinez2006phenomenology}. As regards Shannon's entropy, we can locally measure entropy concerning the states of each cell, its history and its neighbors \cite{lizier2012local}. These entropy measures can be used to numerically calculate the information contained in each cell and its relation to neighbor elements.

These local entropy measures are easy to implement and useful for detecting the existence of a periodic background and the generation of gliders. The first local measure that we can use is the local active information storage (LAIS) to measure how much of the current and past information of a cell defines its future state. For a state $ s \in S $, $LAIS$ is defined as follows:

\begin{equation} \label{eq:LAIS}
LAIS(s)=log_2 \frac{p(s_{n+1} |\mathbf{s}_n^{(k)})}{p(s_{n+1})}
\end{equation}

In Eq. \ref{eq:LAIS} , the conditional probability of having the state $ s $  at the time $ n + 1 $  is reviewed, provided the block $ \mathbf{s}_n^{(k)} $  of $ k $ past states from the time $ k-n + 1$ to the time $ n $. This is divided by the probability of having the state $ s $  at the time $ n + 1 $.

Thus, if a state is defined as the product of the periodic background, the conditional probability $ p(s_{n + 1}) | \mathbf{s}_n^{(k)} $ is expected to be high and, therefore, has a positive $LAIS$ value. On the other hand, if a well-defined background is not present, the conditional probability is low, and its $LAIS$ value tends to be negative. Calculating the $LAIS$ value for each cell in a representative evolution of a cellular automaton, and obtaining its average provides a numerical way of identifying when an automaton is producing information locally from the periodic background.

Another measure of local information quantifies the effect that the information of a neighboring cell has concerning another one and its history. This measure is known as the local transfer entropy ($LTE$). For cellular automata, this measure can be applied to both the right and the left of a given cell.

Given the state $s \in S $ of a cell in a cellular automaton and the state of its right neighbor $ y \in S $, the right $LTE$ of $ s $ can be defined as:

\begin{equation}
LTE_r(s,y)=log_2 \frac{p(s_{n+1} | \mathbf{s}_n^{(k)},y)}{p(s_{n+1} |\mathbf{s}_n^{(k)})}
\end{equation}

The definition of the $LTE$ on the left ($ LTE_l(s, y) $) is similar, although it takes the state  $ y \in S $ of the left neighbor.

The entropy measures $ LTE_r (s, y) $ and $ LTE_l (s, y) $ provide us with an idea of the local information that is transferred between the two neighboring cells of a cellular automaton (either on the left or on the right) in relation to the probability that the state $ s $ is generated by its history. Thus, for the information contained in a glider, $ p (s_{n + 1} | \mathbf{s}_n^{(k)},y) > p (s_{n + 1} | \mathbf{s}_n ^{(k)}) $, which produces positive values of $ LTE_r $ or $ LTE_l $.

As in the case of $LAIS(s) $, the average values of $ LTE_r $ and $ LTE_l $ are calculated numerically through a representative evolution of a cellular automaton, providing us a measure to detect the information transferred between the cells produced by glider interactions.

\section{Complexity, Mean-field Polynomials and Local Information}
\label{complex}

Rule 110 is the prototypical case of a complex cellular automaton. The following table presents the evolution rule, which involves $2$ states and neighborhoods of $3$ cells. 

\begin{table}[th]
	\centering
	\begin{tabular}{|c|c|c|c|c|c|c|c|}
		\hline 
		000 & 001 & 010 & 011 & 100 & 101 & 110 & 111 \\ 
		\hline 
		0 & 1 & 1 & 1 & 0 & 1 & 1 & 0 \\ 
		\hline 
	\end{tabular} 
	\caption{Representation of the evolution rule of Rule 110}
\end{table}

The simulation of Rule 110 with a $ 4 $-state cellular automaton produces the following evolution rule:

\begin{equation}\label{eq:rule110_4states}
\begin{array}{ccc}
\begin{array}{c|cccc}
\multicolumn{1}{c}{}& 00 & 01 & 10 & 11 \\ \cline{2-5}
00 & 00 & 01 & 11 & 11 \\ 
01 & 10 & 11 & 11 & 10 \\ 
10 & 00 & 01 & 11 & 11 \\ 
11 & 10 & 11 & 01 & 00
\end{array} 
&\rightarrow&
\begin{array}{c|cccc}
\multicolumn{1}{c}{}& 1 & 2 & 3 & 4 \\ \cline{2-5}
1 & 1 & 2 & 4 & 4 \\ 
2 & 3 & 4 & 4 & 3 \\ 
3 & 1 & 2 & 4 & 4 \\ 
4 & 3 & 4 & 2 & 1
\end{array} 
\end{array}
\end{equation}

In the first table of Eq. \ref{eq:rule110_4states}, the indexes represent blocks of $ 2 $ binary states by rows and columns, and the input of each table is the evolution of the original rule applied to each block of $ 4 $ states. These blocks have been renamed in the second table to have $ 4 $ states and neighborhoods of size $ 2 $.

\begin{figure}[th]
	\centering
	\includegraphics[width=0.9\linewidth]{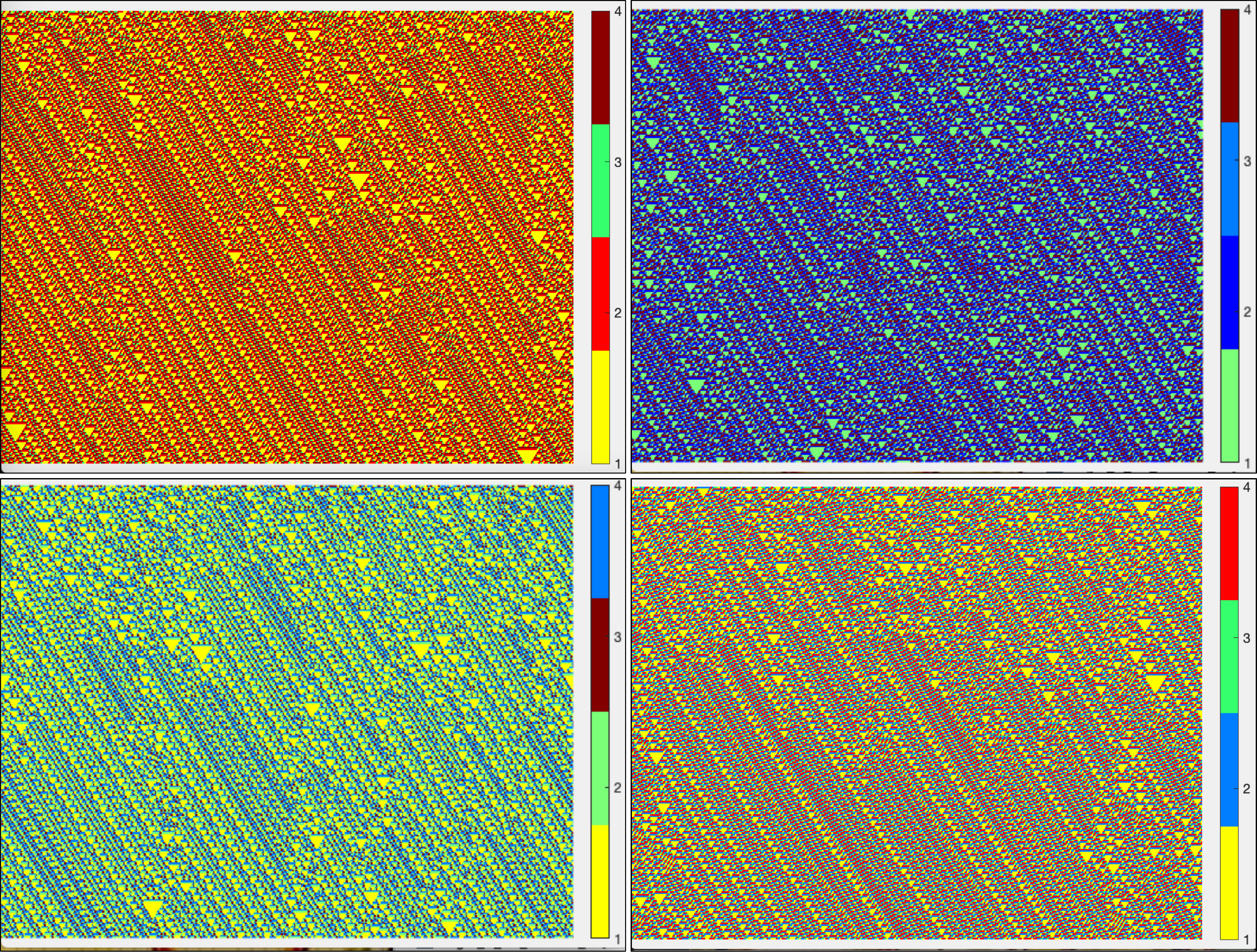}
	\caption{Evolutions of the automaton of $4$ states that simulate Rule 110.}
	\label{fig:ejemplosevolucion1104estados}
\end{figure}

Figure \ref{fig:ejemplosevolucion1104estados} presents examples of the evolution of this automaton by using $ 300 $ cells and $ 300 $ evolutions. Different colors are selected for the $ 4 $ states in each example, and the evolution of every neighborhood has been centered below in the next generation.

To facilitate our analysis, two types of states can be selected: those that shape the background and the additional ones that promote the formation of gliders in this background. Different combinations of states can be chosen in the case of the previous automaton. A combination that works well is taking the states $ 2$ and $4$ as the background. With this, we can filter the evolution of the automaton in only two colors. In these evolutions, state $1$ represents the background states and state $ 2$ the additional ones.

\begin{figure}[th]
\centering
\includegraphics[width=0.9\linewidth]{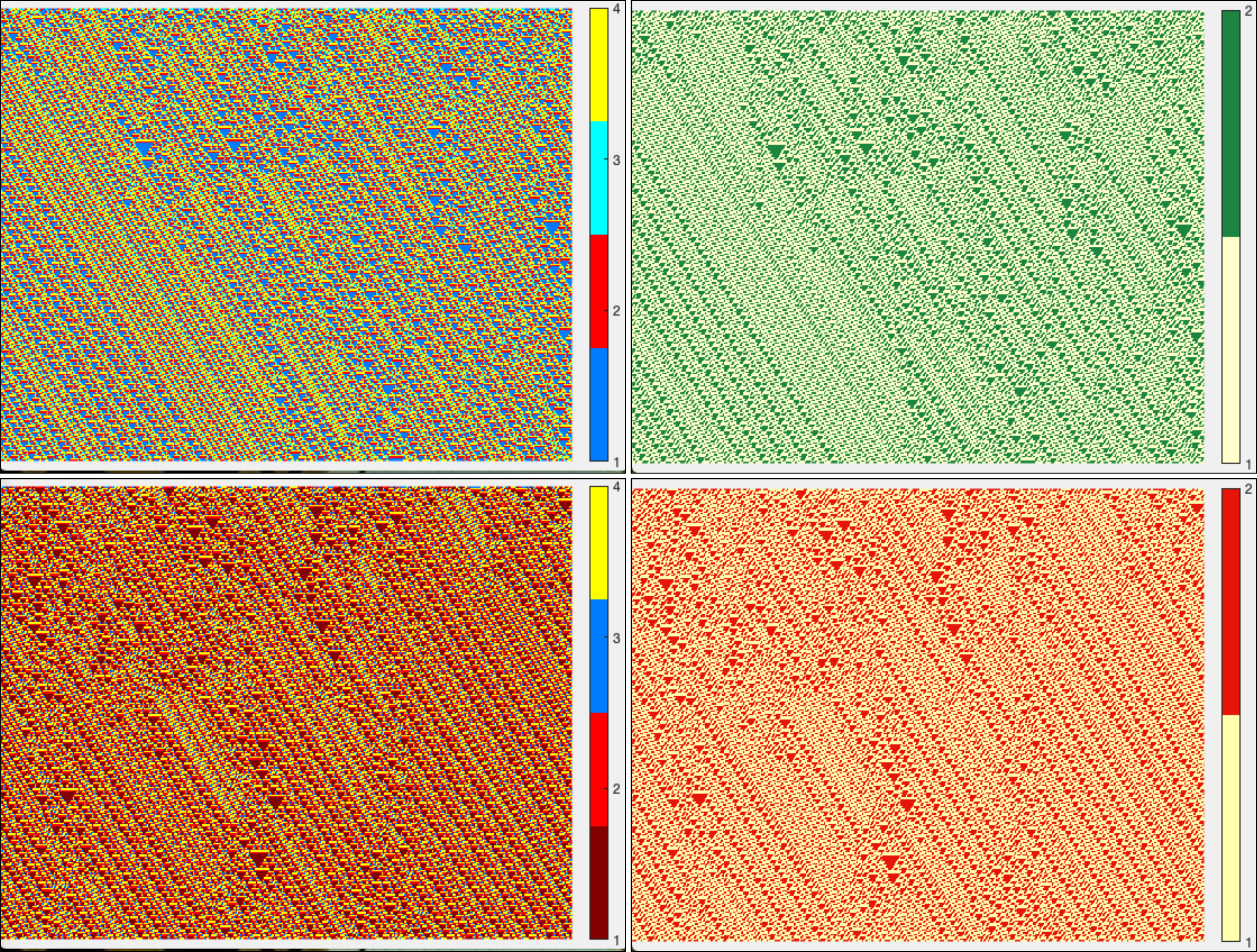}
\caption{Evolution of the automaton of $ 4 $ states (left) applying a filter to differentiate between the background and the gliders (right).}
\label{fig:ejemplosevolucion1104estados_filtro}
\end{figure}

\begin{figure}[th]
	\centering
	\includegraphics[width=1\linewidth]{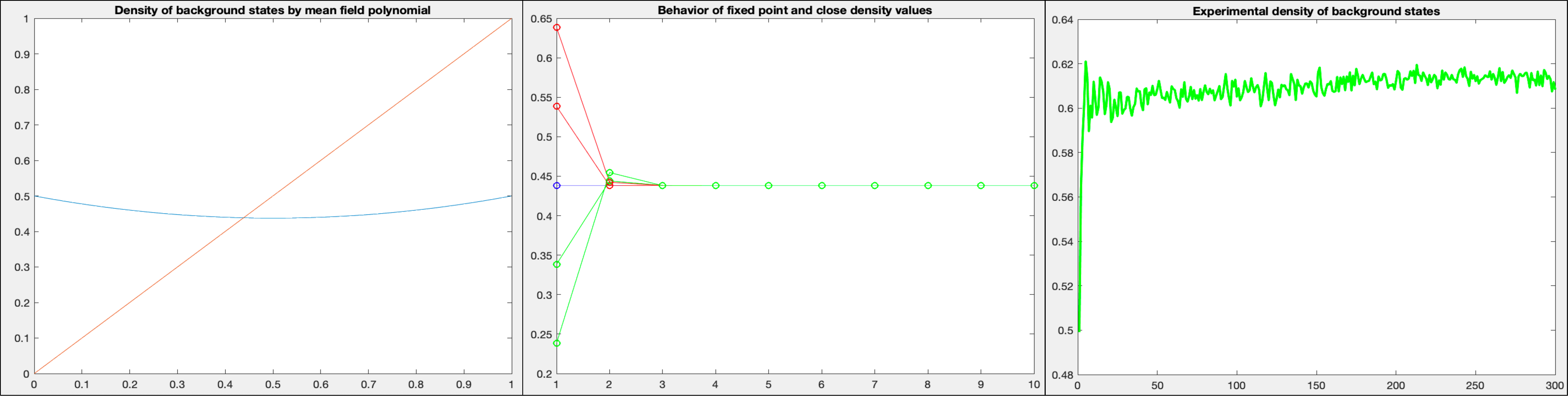}
	\caption{Density estimated by the mean-field polynomial, fixed-point behavior and close values in $10$ iterations of the polynomial, and experimental density of background states.}
	\label{fig:ejemplosevolucion1104estados_CampoPromedio}
\end{figure}

Let $ q $ be the density of background states for this rule. The mean-field polynomial for states $ 2$ and $4 $ is:

\begin{equation} \label{eq:mean_field_4states}
\begin{array}{l}
q'=\frac{1}{2}q^2+\frac{3}{4}q(1-q)+\frac{1}{2}(1-q)^2 \\ \\
q'=\frac{1}{2}q^2+\frac{3}{4}q-\frac{3}{4}q^2+\frac{1}{2}-q+\frac{1}{2}q^2 \\ \\
q'=\frac{1}{4}q^2-\frac{1}{4}q+\frac{1}{2}
\end{array}
\end{equation}

where $q'$ is the resultant density, given an initial $q$. Figure \ref{fig:ejemplosevolucion1104estados_CampoPromedio} presents the graph of the mean-field polynomial of the background states against the identity of the density so that the fixed points can be appreciated. Moreover, it shows the experimentally measured density of these states in $ 30 $ evolution samples, each with $ 300 $ states and $ 300 $ evolutions.

The mean-field polynomial estimates a stable density at a value of $0.4348$, while the sample has an experimental density of around $0.61$. The polynomial has a discrepancy because it is approaching two states at the same time; therefore, its prediction falls below the real value. However, the critical point is that the fixed value it predicts is stable, which can be observed in the graph and the polynomial iteration.

The polynomial suggests that the density of background states is preserved without being too high or low (i.e., around $ 0.5 $), which means that the rest of the additional states are also preserved and can form structures in a periodic background.

A stable density of background states that is not too dominant is a crucial point to allow the formation of gliders. We can also use the local information and entropy measures to characterize the composition of gliders in the evolution of an automaton.

Let us take a random $ 4 $ state automaton with the following evolution rule:

$$
M_\varphi=
\begin{array}{c|cccc}
\multicolumn{1}{c}{}& 1 & 2 & 3 & 4 \\ \cline{2-5}
1 & 4 & 1 & 3 & 1 \\ 
2 & 4 & 4 & 2 & 4 \\ 
3 & 2 & 3 & 1 & 3 \\ 
4 & 3 & 2 & 3 & 3
\end{array} 
$$

\begin{figure}[th]
	\centering
	\includegraphics[width=1\linewidth]{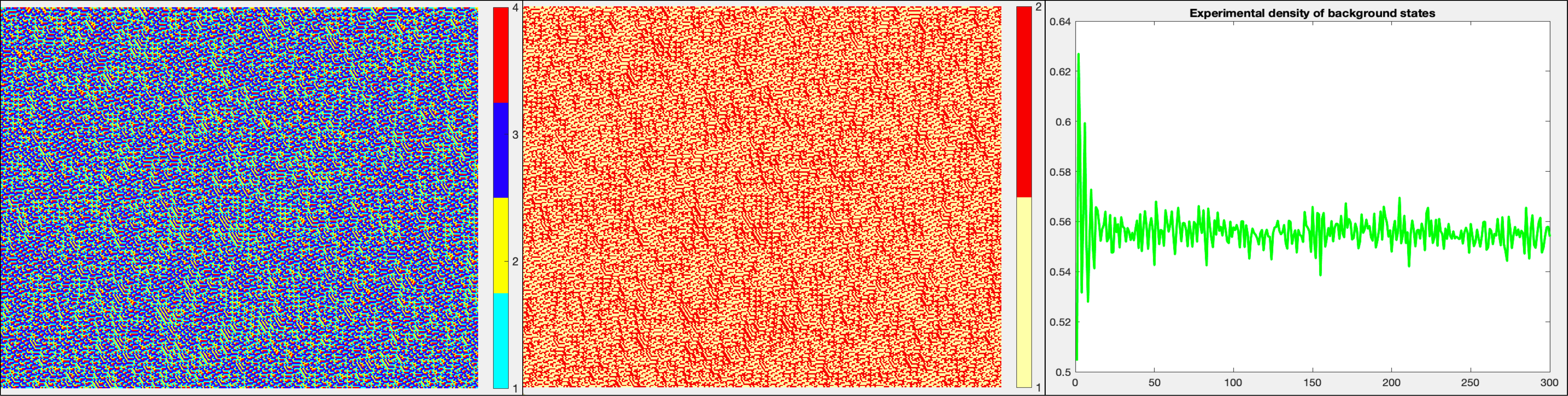}
	\caption{Regular evolution with filter and density of background states in a $ 4 $- state automaton.}
	\label{fig:EjemploEvolucion_NormalFiltroDensidad_ACAleatorio_4estados}
\end{figure}

Fig. \ref{fig:EjemploEvolucion_NormalFiltroDensidad_ACAleatorio_4estados} presents the examples of its regular and filter evolution by selecting the states $ 3$ and $ 4$ as the  background and taking $ 300 $ cells and $ 300 $ evolutions as well as the density of this filter with the average of $ 30 $ samples.

In this case, the background states tend to retain a density close to $0.56$ as the automaton evolves. However, the presence of structures interacting in the evolution space is not observed as in complex automata. We can use local entropy measures to characterize the information features in this automaton. The analysis of local entropies is carried out on the filtered evolution to facilitate our study. Thus, we only consider two types of states: background and additional states, which make the computation of entropies easier. For the analysis, we take a sample filtered evolution of $ 10,000 $ cells with random states and $ 600 $ evolutions to approximate the values of $LAIS$, $LTE_r $ and $LTE_l $. Then, those values are used to make the measurements in the blocks of states obtained in another evolution and obtain an average value.

\begin{figure}[th]
	\centering
	\includegraphics[width=1\linewidth]{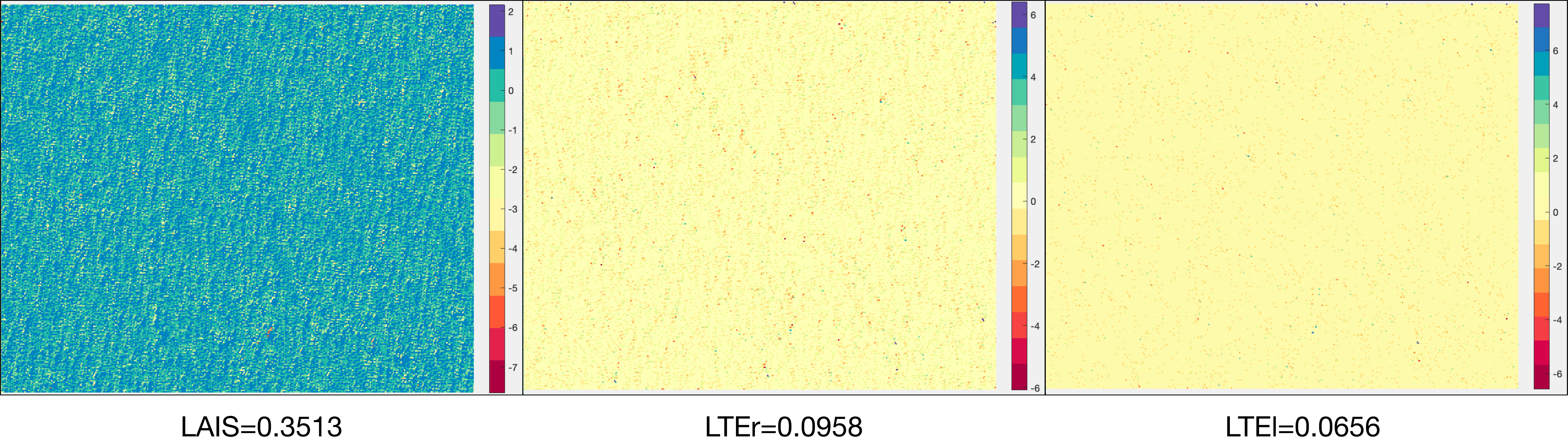}
	\caption{Local active information storage (LAIS) and local entropies $LTE_r$ and $LTE_l$ in a $4$-state automaton with filtered evolution.}
	\label{fig:EjemploEntropiasLocales_ACAleatorio_4estados}
\end{figure}

In this example, we observe that the value of $LAIS$ is $ 0.3513$, which, relative to the local information that holds the background states, is a low value. On the other hand, the values of the information that is maintained by the additional states (in this case, the $1$ and $2$ states) are, on average, disperse with $LTE_r = 0.0958 $ and $LT _l =  0.0656$.

Let us now conduct the same analysis for a random evolution of the $ 4$- state cellular automaton that emulates Rule 110 by taking a filtered evolution of $ 300 $ cells and $ 300 $ evolutions.

\begin{figure}[th]
	\centering
	\includegraphics[width=1\linewidth]{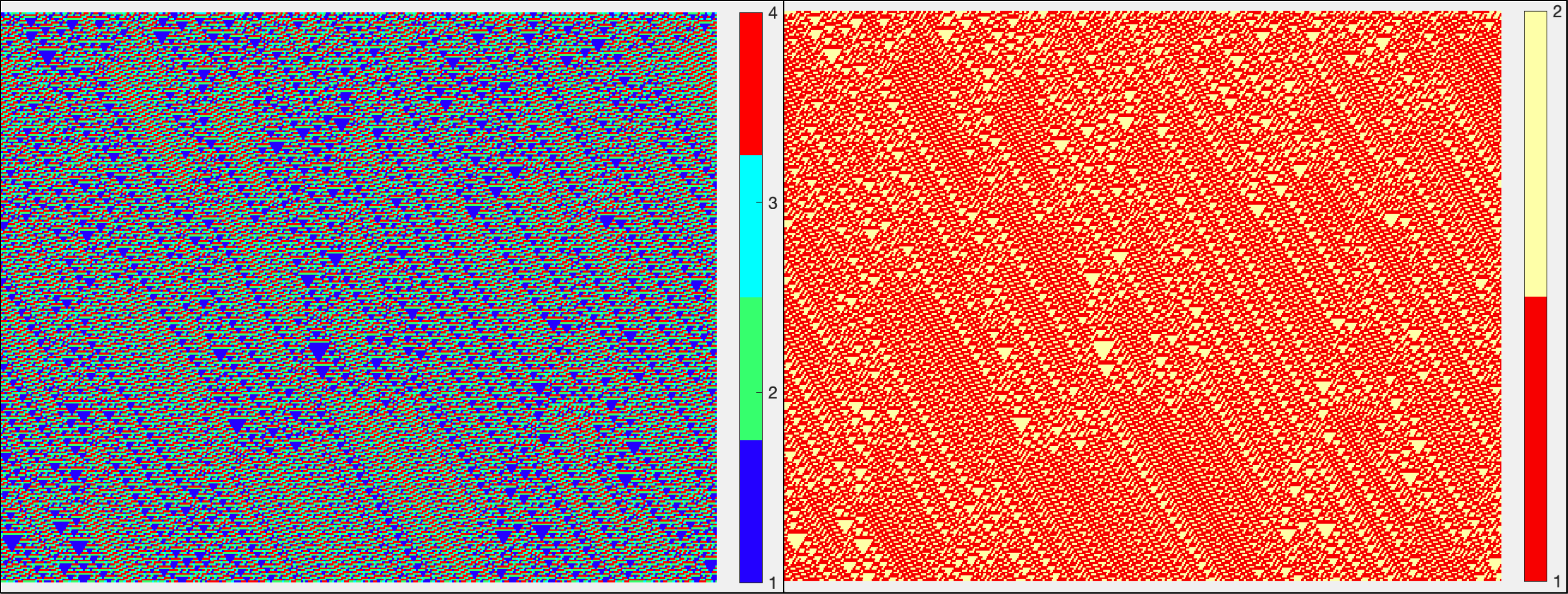}
	\caption{Regular and filtered evolution of the $4$-state automaton emulating Rule 110.}
	\label{fig:EjemploEvolucion_NormalFiltro_AC110_4estados}
\end{figure}

\begin{figure}[th]
	\centering
	\includegraphics[width=1\linewidth]{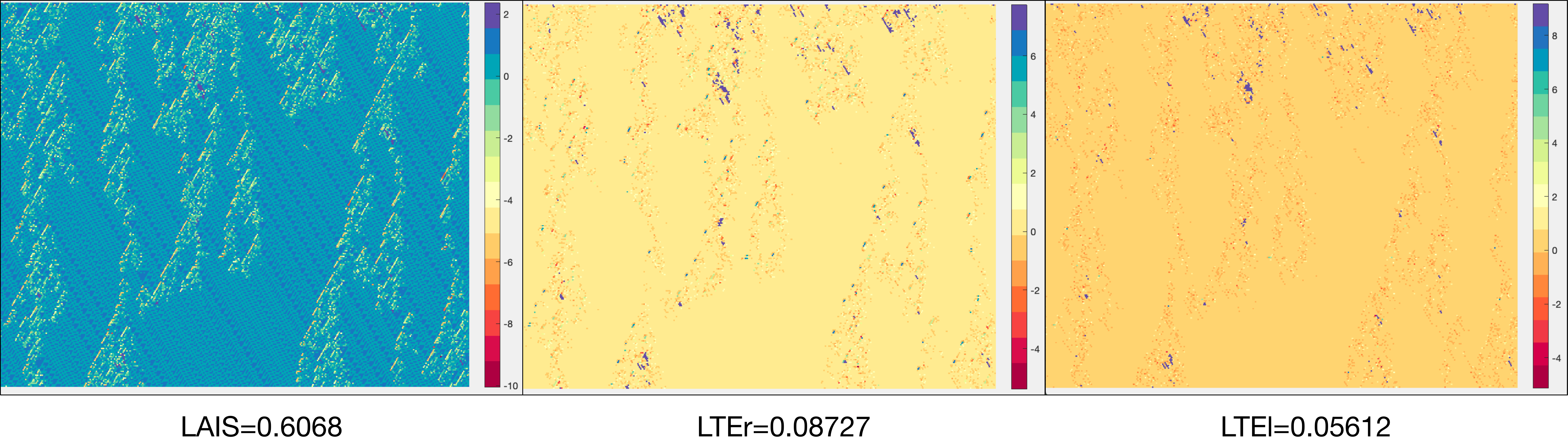}
	\caption{Local active information storage (LAIS) and local entropies $LTE_r$ and $LTE_l$ in the filtered evolution of the $4$-state cellular automaton emulating Rule 110.}
	\label{fig:EjemploEntropiasLocales_ACRegla110_4estados}
\end{figure}

We know that the background states present a density close to $ 0.6 $ as the automaton evolves. Figure \ref{fig:EjemploEntropiasLocales_ACRegla110_4estados} shows the measures of $LAIS$, $LTE_r $ and $LTE_l$ for this evolution.  First, $ 10,000 $ cells with random states and $ 600 $ evolutions are taken to approximate the local information values. Then, these measurements are applied in the sample evolution with filter, calculating their averages with the obtained blocks of the states.

Figure \ref{fig:EjemploEntropiasLocales_ACRegla110_4estados} shows that the local information retained in the background states is higher than the last example ($LAIS = 0.6068 $) and, although the local transfer entropy values are on average similar to the previous example, they, unlike this, are not homogeneously dispersed, indicating the interaction of structures in a periodic background.

This study uses the same analysis for the generation of complex automatons with a higher number of states looking for a conservation of the density of background states close to $ 0.6 $, an average $LAIS$ value close to $0.6 $ and an average value of the local transfer entropies similar to those observed for the $ 4 $-state cellular automaton emulating Rule 110.

\section{Random Generation of Complex Cellular Automata with Multiple States}

The purpose of this study is to demonstrate that it is possible to generate complex cellular automata with hundreds of states. For this, first, a cellular automaton is randomly generated with $a$ states. This automaton serves as a periodic background. Then, the evolution rule of this first automaton is expanded with $b$ additional states so that we obtain a newly expanded automaton of $a+b$ states. The neighborhoods resulting from the combination of the $a$ initial states with the additional $b$ states are filled at random with uniform probability after taking the whole of $a+b$ states into account.

Formally, the process can be defined as follows: To begin with, a  cellular automaton $\mathbb{A} = \{S, \varphi, 2 \} \} $ is randomly defined such that:

\begin{equation}
M_\varphi(s,y)=rand(S) \;\; \forall s,y \in S
\end{equation}

where $rand(S)$ is a random state of $S$. Once $\mathbb{A}$ is defined, another automaton $\mathbb{P}=\{ S \cup P, \varphi^\prime, 2 \}$ is defined based on it, where $P$ is a new automaton with a set of states such that:

\begin{equation}
M_{\varphi^\prime}(s,y) = \left\{ \begin{array}{l} M_\varphi  \iff s,y \in S \\ \\ rand(S \cup P)  \iff s \in P  \; \lor \;  y \in P \end{array}  \right.
\end{equation}

Notably, the evolution rule $M_{\varphi^\prime}$ is an extension of $M_{\varphi}$ that preserves the original part for neighborhoods formed with the states of $S$ but takes any random state in $S  \cup P$ for the rest of the new neighborhoods. This definition of $\mathbb{P}$ allows the original automaton $\mathbb{A}$ to act as a periodic background because it is a closed subsystem in the evolution of $\mathbb{P}$ and for which some of their new neighborhoods produce background states randomly.

Given the cellular automaton $\mathbb{P}= \{S \cup P, \varphi^\prime, 2 \} $, with $ a = | S | $ and $ b = | P | $, there are $ (a + b)^2 = a^2 + 2ab + b^2 $ neighborhoods. We know that $ a^2 $ neighborhoods produce only background states, while $ 2ab + b^2 $ neighborhoods randomly generate both background and additional states. Thus, due to uniform randomness, the probability of generating any state in these $ 2ab + b^2 $ neighborhoods is:

$$ \frac{1}{a+b} $$.

Hence, the probability of obtaining background states in the new neighborhoods is:

$$
\frac{a}{a+b}
$$

Let $ q $ be the probability of having a background state in a given evolution of the $\mathbb{P}$ automaton. Then, the probability of having a neighborhood formed by the background states is $q^2 $, the probability of having neighborhoods formed by additional states is $ (1-q)^2 = 1-2q + q^2 $ and the probability of having mixed neighborhoods is $2q(1-q) = 2q-2q^2 $. The proportion of neighborhoods formed with background states that evolve in background states is $1$, and the proportion of the other types of neighborhoods that generate background states is $\frac{a}{a + b}$. In this way, the mean-field polynomial that approximates the density $ q '$ of background states in the following evolution can be defined as:

\begin{equation} \label{eq:densityBackgroundStates}
\begin{array}{lll}
q'& = & q^2 + \left( \frac{a}{a+b} \right) \left(2q-2q^2\right) + \left( \frac{a}{a+b} \right) \left(1-2q+q^2\right)  \\
&& \\
& = & \left(1- \frac{a}{a+b} \right)q^2+\left( \frac{a}{a+b} \right)
\end{array}
\end{equation}

\begin{figure}[th]
	\centering
	\includegraphics[width=1\linewidth]{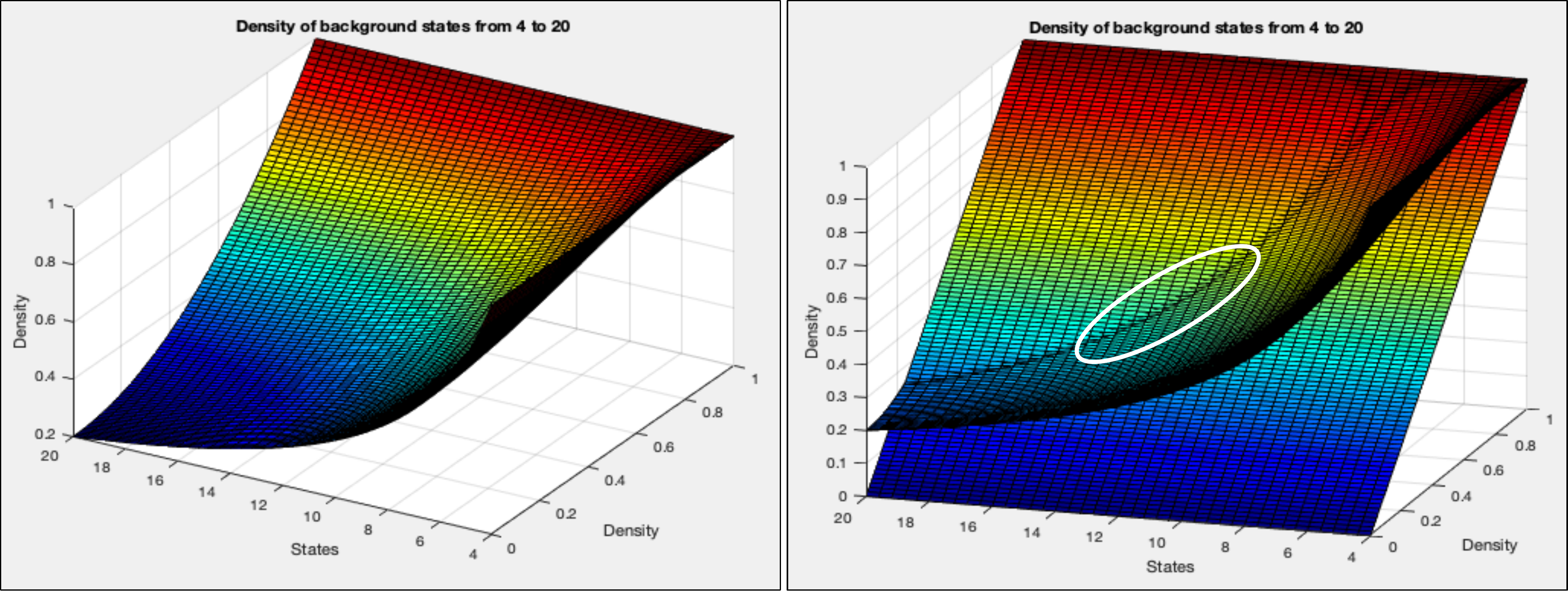}
	\caption{(A) Surface representing the density of the background states approximated by mean-field polynomial. (B) The same surface against the identity; the desired densities are represented by the white ellipse.}
	\label{fig:EjemploDensidad_MultiplesEstados}
\end{figure}

Figure \ref{fig:EjemploDensidad_MultiplesEstados} shows the surface obtained by iterating the polynomial of the previous equation for all the densities, with $ a = 4 $ and $ 0 \leq b \leq 16 $. In the section on the right, we observe the surface crossed with the surface that represents the identity of the densities. The intersection exemplifies the densities that are fixed points in Eq. \ref{eq:densityBackgroundStates}.

We are interested in obtaining the densities of the background states close to $ 0.6$, remembering that the mean-field polynomial is underestimating the density. Therefore, we are searching for a behavior owing to which there is a balance between the background states and the additional states for the formation of gliders in a periodic background.

In this manner, we can extend the cellular automaton of $ 4 $ states to one between $ 10 $ and $ 14 $ states in order to obtain a density that allows the formation of gliders.

To take an approximation of the number of additional states needed to obtain complex behaviors, let us assume $ b $ is proportional to $ a $, i.e., $ b = \alpha a $. Then, Eq. \ref{eq:densityBackgroundStates} can be expressed as:

\begin{equation} \label{eq:densityBackgroundStates-2}
q'=\left(1- \frac{a}{a+\alpha a} \right)q^2+\left( \frac{a}{a+ \alpha a} \right)
\end{equation}

Now, suppose we want to find the $ \alpha $ ratio so that the mean-field equation preserves the density of the background states. We now have the following equation:

\begin{equation} \label{eq:densityBackgroundStates-3}
q=\left(1- \frac{a}{a+\alpha a} \right)q^2+\left( \frac{a}{a+ \alpha a} \right)
\end{equation}

By solving $ \alpha $, we establish Eq. \ref{eq:densityBackgroundStates-4}:

\begin{equation} \label{eq:densityBackgroundStates-4}
\begin{array}{rcl}
q &=& q^2-\left( \frac{a}{a+ \alpha a} \right)q^2+\left( \frac{a}{a+ \alpha a} \right) \\
&& \\
(a+ \alpha a)q &=& (a+ \alpha a)q^2-aq^2+a \\
&& \\
aq+ \alpha aq &=& \alpha a q^2 +a \\
&& \\
\alpha q - \alpha  q^2 &=& 1 - q \\
&& \\
\alpha &=&\frac{1 - q}{q -  q^2} = \frac{1 - q}{q(1-q)} = \frac{1}{q} \\
\end{array}
\end{equation}

The proportion of additional states concerning the number of background states can be approximated by inversing the density of background states that one wishes to have in the evolution of the automaton.

This proportion indicates that we should look for the number of appropriate additional stages in order to obtain the complex behaviors that use a random cellular automaton as a periodic background.

Let us take a cellular automaton $ \mathcal{A} = \{4, \varphi, 2 \} $ defined randomly as:


\begin{figure}[th]
	\centering
	\includegraphics[width=0.75\linewidth]{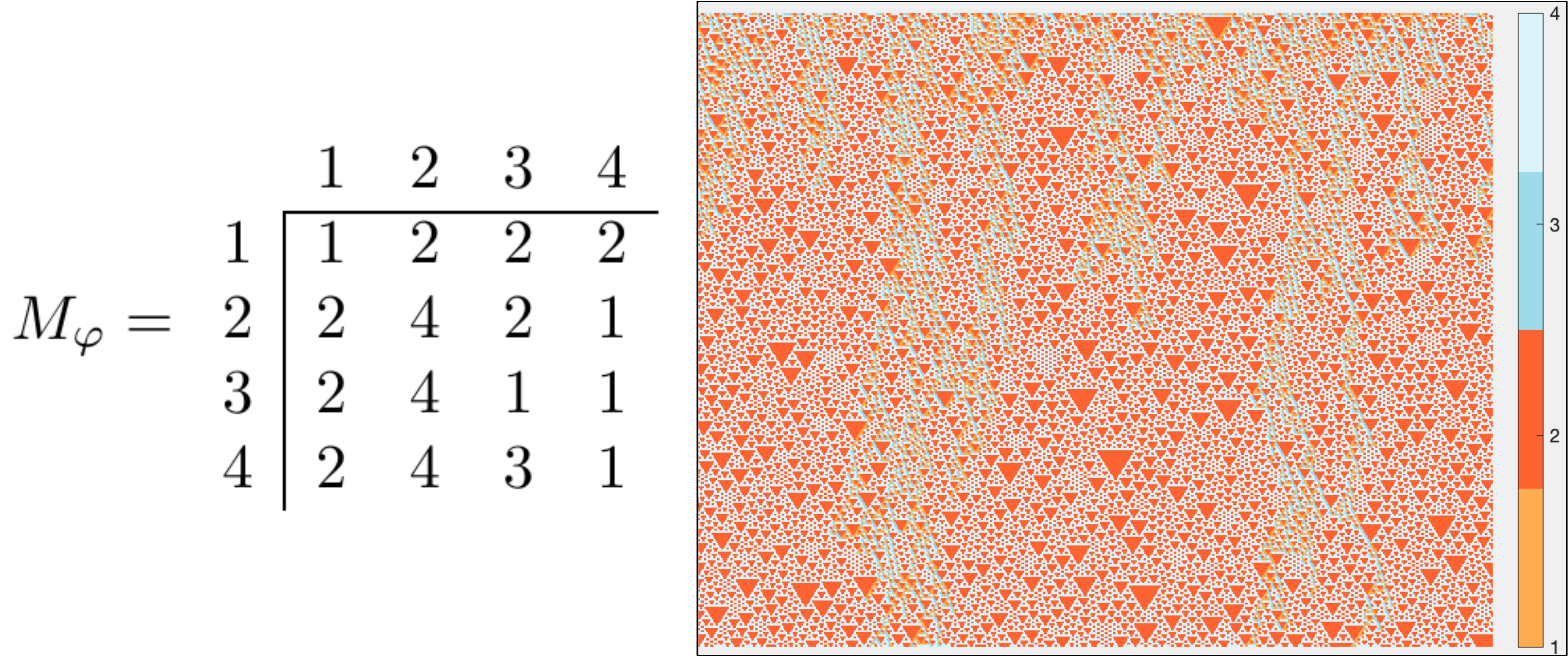}
	\caption{Random cellular automaton of $4$ states}
	\label{fig:Ejemplo_Regla4hAleatoria}
\end{figure}

Let us take this automaton as the basis for extending it to an automaton with a complex behavior. Eq. \ref{eq:densityBackgroundStates-4} tells us that if we want a density of $ q = 0.5 $ for the background states, we must add $ \alpha 4 $ more states with $ \alpha = 1 / 0.5 = 2 $. Thus, we extend our rule to obtain a new automaton $ \mathcal{P} = \{12, \varphi', 2 \} $.

\begin{figure}[th]
	\centering
	\includegraphics[width=0.75\linewidth]{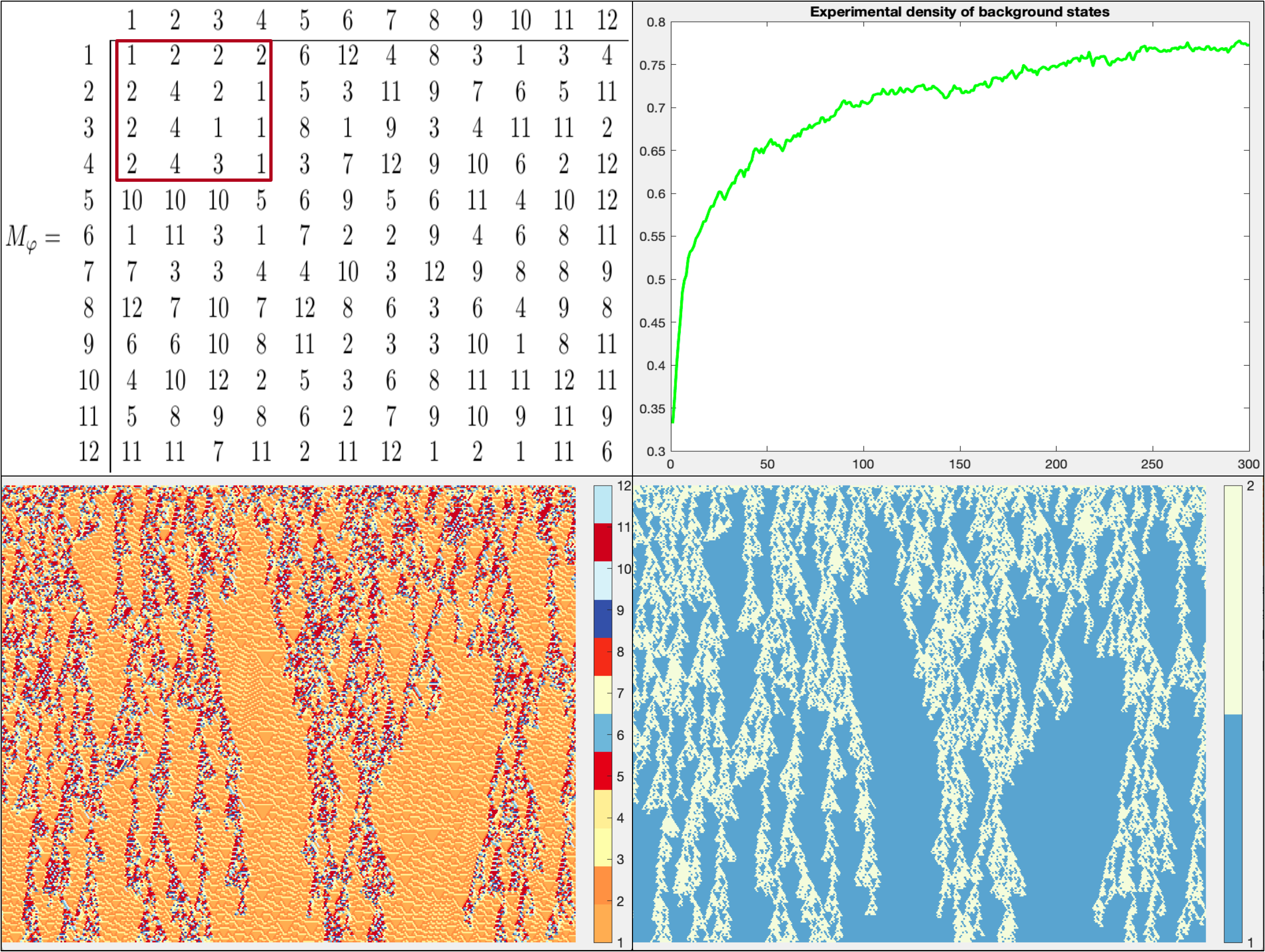}
	\caption{Randomly extended cellular automaton based on a $ 4 $-state automaton.}
	\label{fig:Regla4h01_Extendida12Estados}
\end{figure}

Figure \ref{fig:Regla4h01_Extendida12Estados} shows that the extended evolution rule has $ 12 $ states compared with the original $ 4 $ (also indicated in the figure). Moreover, it depicts a sample and a filtered evolution, demonstrating the background states (from $ 1$ to $ 4$) in one color and the additional states in another. Furthermore, the experimental density of the background states is close to an average of $ 0.75 $, which is above the $ 0.5 $ defined in the Equation \ref{eq:densityBackgroundStates-4}. Although this equation is not exact, it does provide us with a clue for defining a suitable number of states in order to find complex behaviors.

Let us consider the filtered evolution. The local information and transfer entropies are estimated as follows:

\begin{figure}[th]
	\centering
	\includegraphics[width=0.95\linewidth]{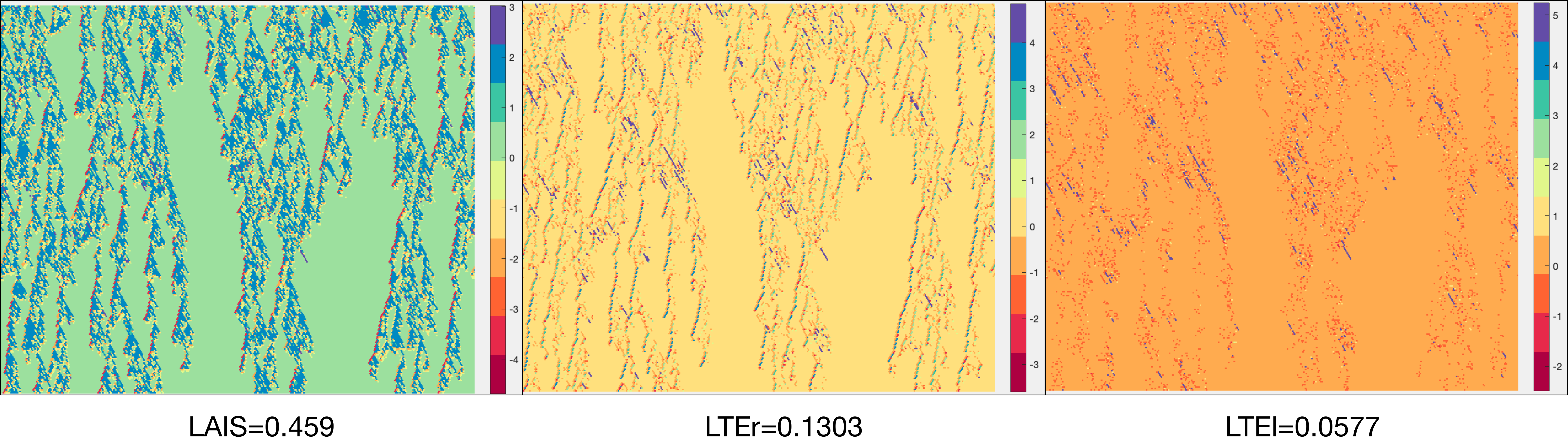}
	\caption{Local information and transfer entropies for the extended cellular automaton of $12$ states}
	\label{fig:Regla4h01_Extendida12Estados_Entropias}
\end{figure}

The value of $LAIS$ is lower compared to the $4$-state cellular automaton that emulates Rule 110 but with its $LTE_r$ value is higher and $LET_l$ value is similar. Let us consider another cellular automaton $\mathcal{A}=\{4, \varphi, 2 \}$, which is also randomly defined as:


\begin{figure}[th]
	\centering
	\includegraphics[width=0.75\linewidth]{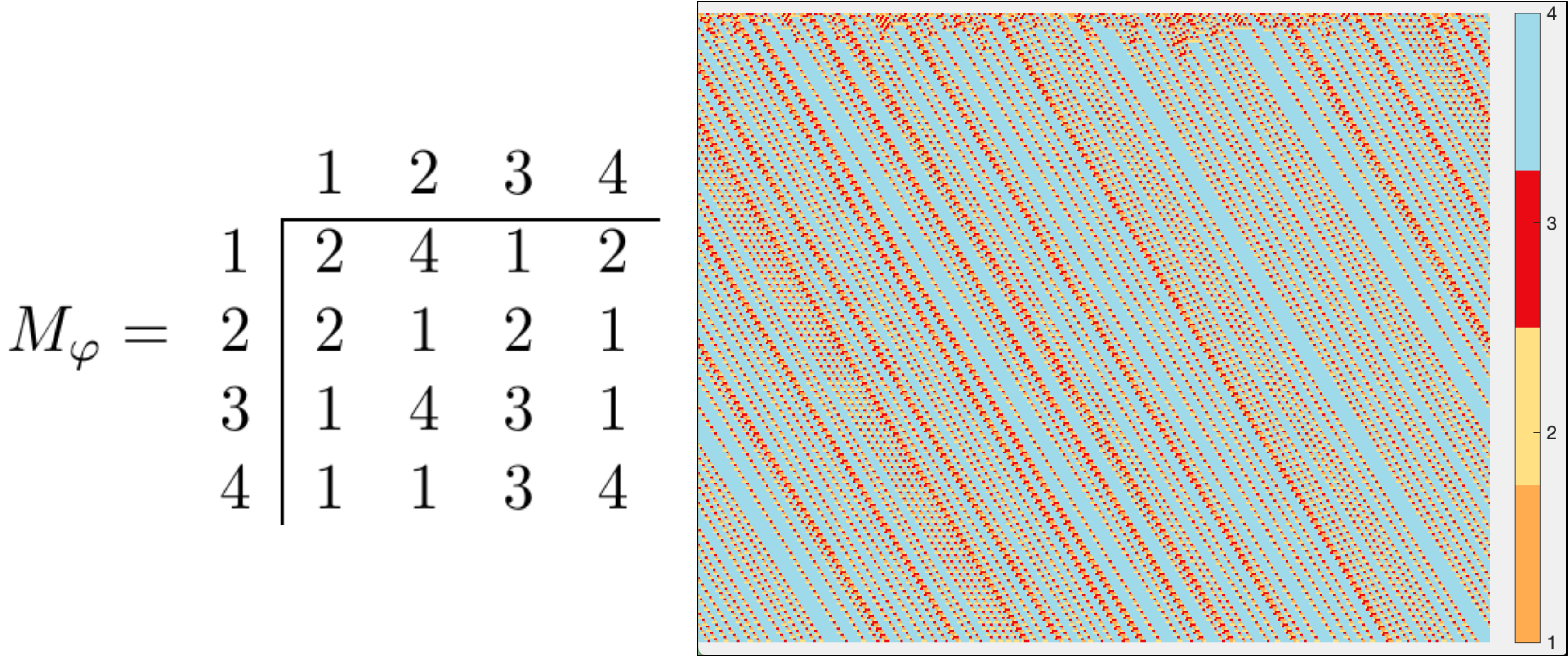}
	\caption{Random cellular automata of $4$ states.}
	\label{fig:Ejemplo_Regla4hAleatoria2}
\end{figure}

The previous automaton has a periodic behavior in a few evolutions. We take this automaton as the basis for extending it to an automaton with a complex behavior but add more states in it than in the previous one. Thus, we expand our rule to obtain a new cellular automaton $ \mathcal{P} = \{14, \varphi ', 2 \} $.



\begin{figure}[th]
	\centering
	\includegraphics[width=0.75\linewidth]{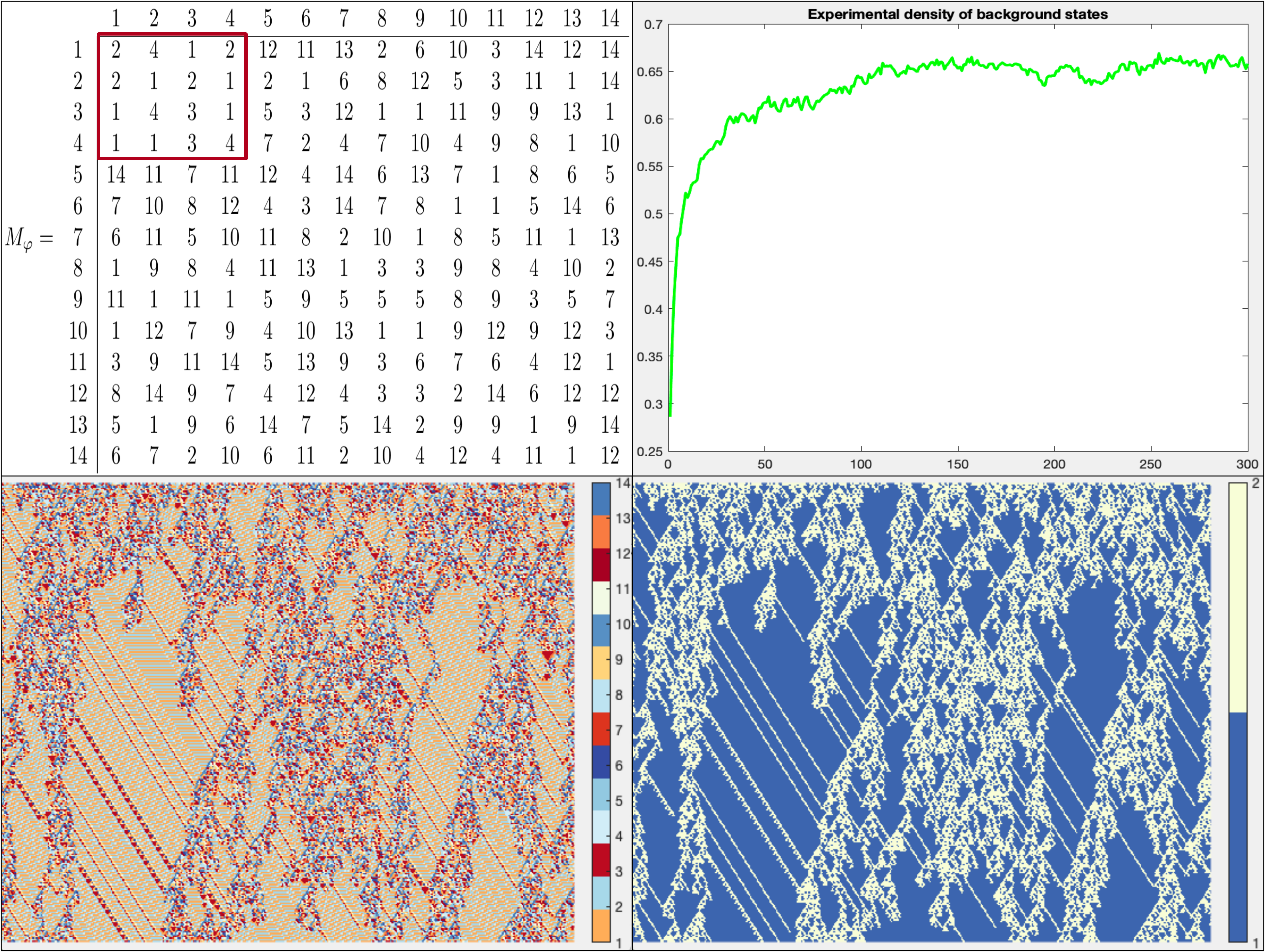}
	\caption{Cellular automaton of $14$ states extended randomly from a $4$-state automaton.}
	\label{fig:Regla4h02_Extendida14Estados}
\end{figure}

Figure \ref{fig:Regla4h02_Extendida14Estados} presents the extended evolution rule of $ 14 $  states along with a sample and the filtered evolution. The value of the experimental density of the background states is close to an average of $ 0.647$, which is closer to the $ 4 $-state automaton that emulates Rule 110. In this case, the filtered evolution, local information and transfer entropies are estimated as follows:

\begin{figure}[th]
	\centering
	\includegraphics[width=0.95\linewidth]{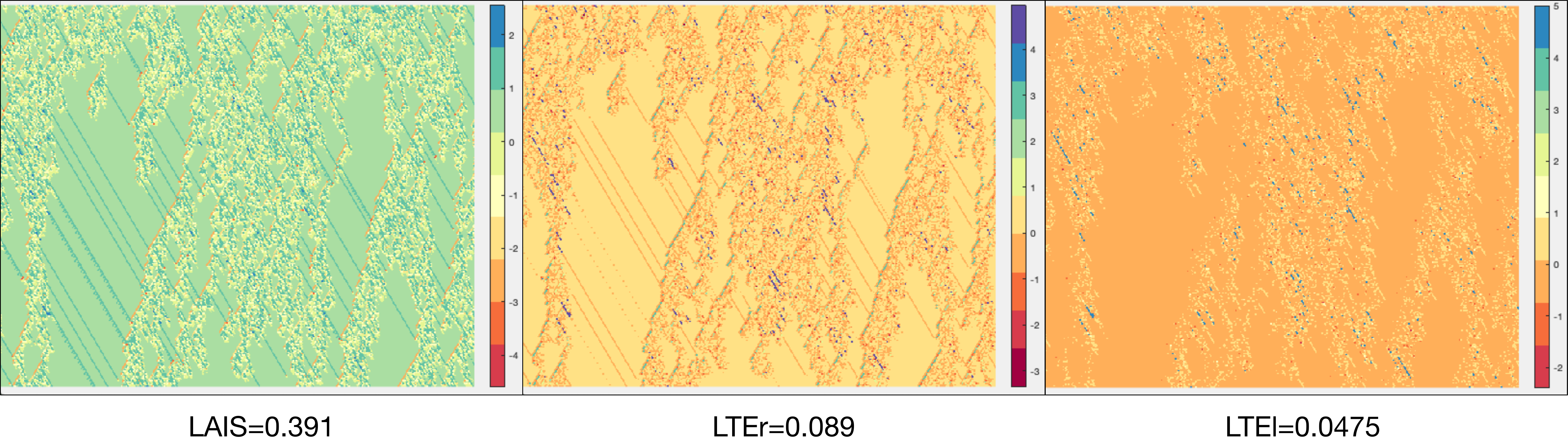}
	\caption{Local information and transfer entropies for the extended cellular automaton of $14$ states.}
	\label{fig:Regla4h01_Extendida14Estados_Entropias}
\end{figure}

Once again, the value of $LAIS$ is lower compared to the $4$-state cellular automaton that simulates Rule 110. Moreover, although its $LTE_r$ value is similar, its  $LET_l$ value is again a little lower. The density equation (Eq. \ref{eq:densityBackgroundStates-4}) and the experiment indicates that we must look for an $\alpha$ ratio between $ 2 $ and $ 2.5 $ additional states in order to find complex behaviors through a random cellular automaton with $a$ initial states.

\section{Obtaining Complex Cellular Automata with Multiple States}


Given a random cellular automaton with $a$ states, the previous results provide us with a set of parameters to generate complex cellular automata. We look for extended automata with $ 3a $ to $ 3.5a $ states, with the $ LAIS $ value greater than $ 0.4 $ and the transfer entropies $ LTE_r $ and $ LTE_i $ greater than $ 0.05$.

Of course, not all the extensions of a cellular automaton generates complex behaviors. Therefore, a simple genetic algorithm is applied to improve the evolution rules so that they show increasingly complex behaviors and make this search more efficient.

The genetic algorithm proposed in this study takes $ 30 $ individuals (evolution rules) in the matrix form and $ 600 $ iterations. Initially, each individual shares the same original rule of $ a$ states and their new neighborhoods are defined uniformly and randomly among all the possible states.
In each iteration, individuals are ranked after considering their local information and transfer entropy values (weighted equally). Based on this qualification, a refined population is selected by using a tournament strategy. With this population improved, each rule is crossed with another randomly selected rule, with a crossing probability of $ 0.5.$ Then, each modified rule is mutated only in the neighborhoods of the states that contain at least one additional element, with a probability of $ 0.1 $. If this modified rule improves on the original one, it takes its place; otherwise, the original rule remains.
 
This simple genetic algorithm calculates the cellular automata with complex behaviors with $ 2$, $ 8$, $ 16$ and $ 32 $ initial states that are used as the background and extend later to the $ 7$, $28$ , $56$ and $112$ states, respectively. 

 \begin{figure}[th]
 	\centering
 	\includegraphics[width=0.75\linewidth]{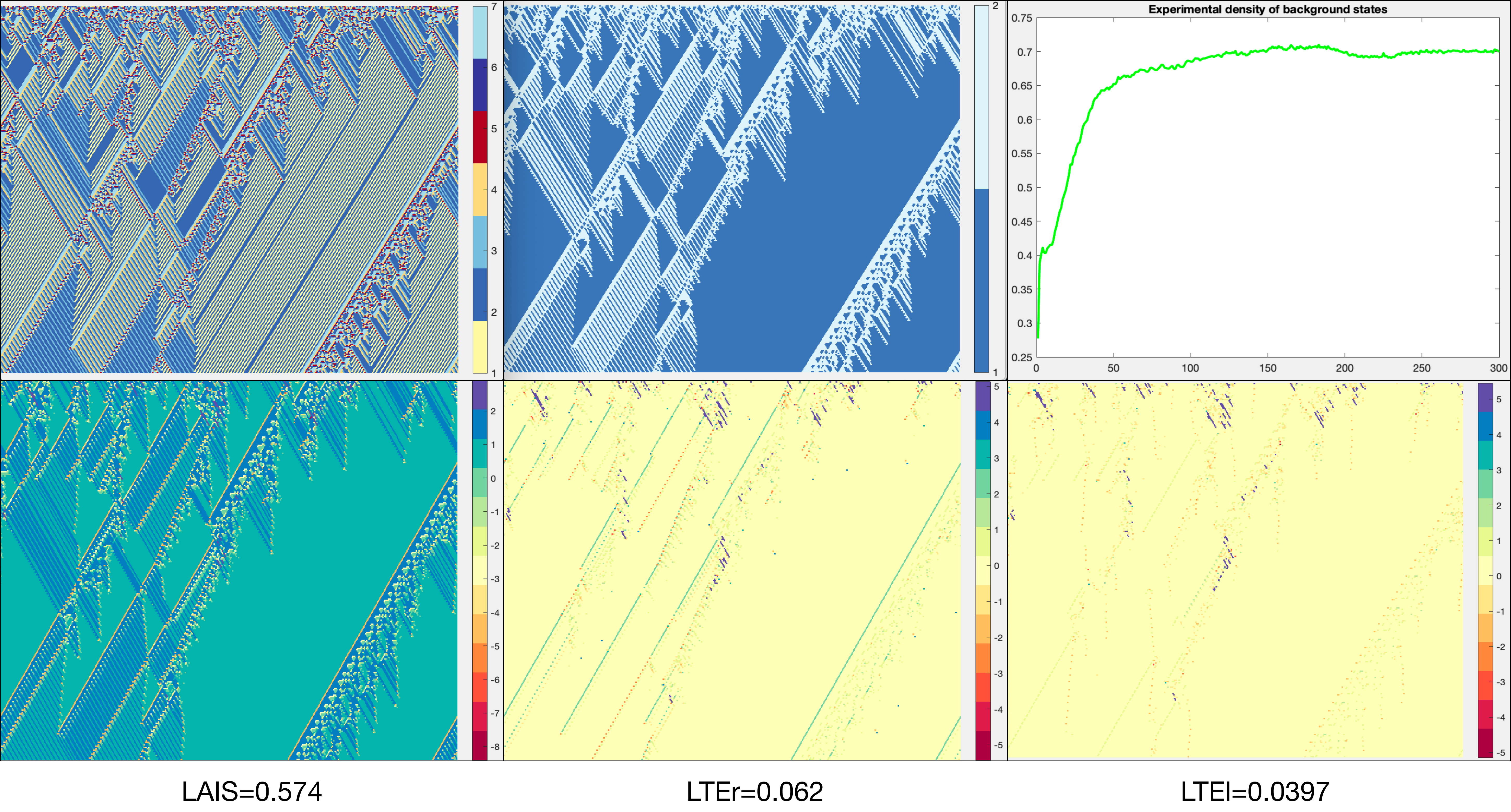}
 	\caption{Randomly extended cellular automaton of $7$ states from a cellular automaton of $2$ states.}
 	\label{fig:AutomataComplejo_7Estados}
 \end{figure}
 
The cellular automaton in Figure \ref{fig:AutomataComplejo_7Estados} uses as a random automaton of $ a=2 $ states as a basis and then extends with a proportion of  $\alpha= 2.5 $. Hence, we finally have  $a+\alpha a =7 $ states . By filtering the evolution in the background and additional states, the movement of gliders in a periodic background can be appreciated. The experimental tests demonstrate that the density of the background states is close to $ 0.7$, while the local entropy values are close to those demonstrated by the emulation of Rule 110 with $ 4 $ states.

\newpage

 \begin{figure}[th]
	\centering
	\includegraphics[width=0.75\linewidth]{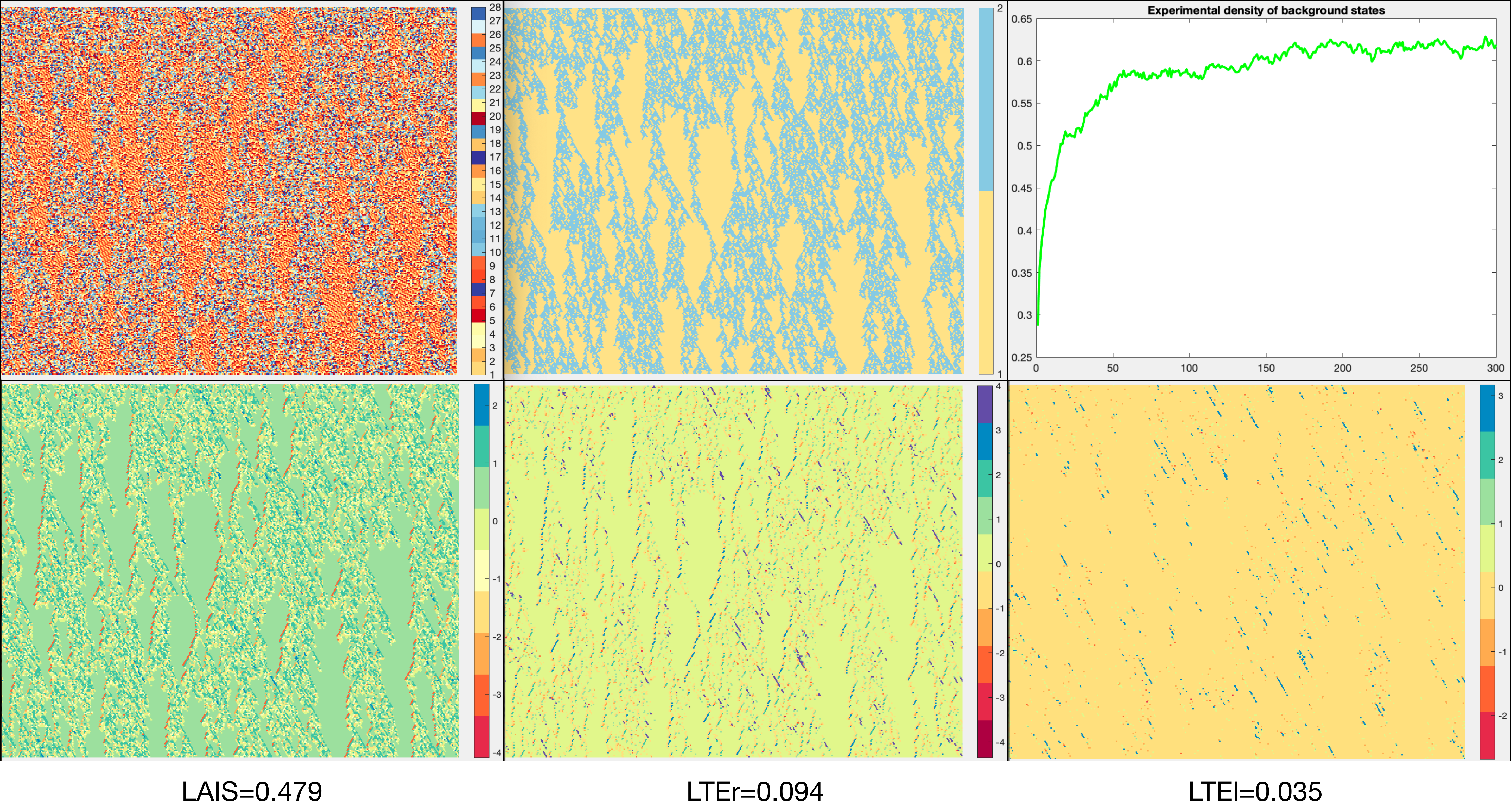}
	\caption{Randomly extended cellular automaton of $28$ states using a random cellular automaton of $8$ states as background.}
	\label{fig:AutomataComplejo_28Estados}
\end{figure}

The cellular automaton in Figure  \ref{fig:AutomataComplejo_28Estados} takes a random automaton with $ 8 $ states as the background. In the initial evolution, it is possible to perceive the emergence of structures in a multicolored periodic background. The filter offers a better perception of these structures by differentiating between the background states and the additional states. Its experimental density is close to $ 0.62$, while the values of the local information and transfer entropies are close to those corresponding to Rule 110 simulated with $ 4 $ states.

\begin{figure}[th]
	\centering
	\includegraphics[width=0.75\linewidth]{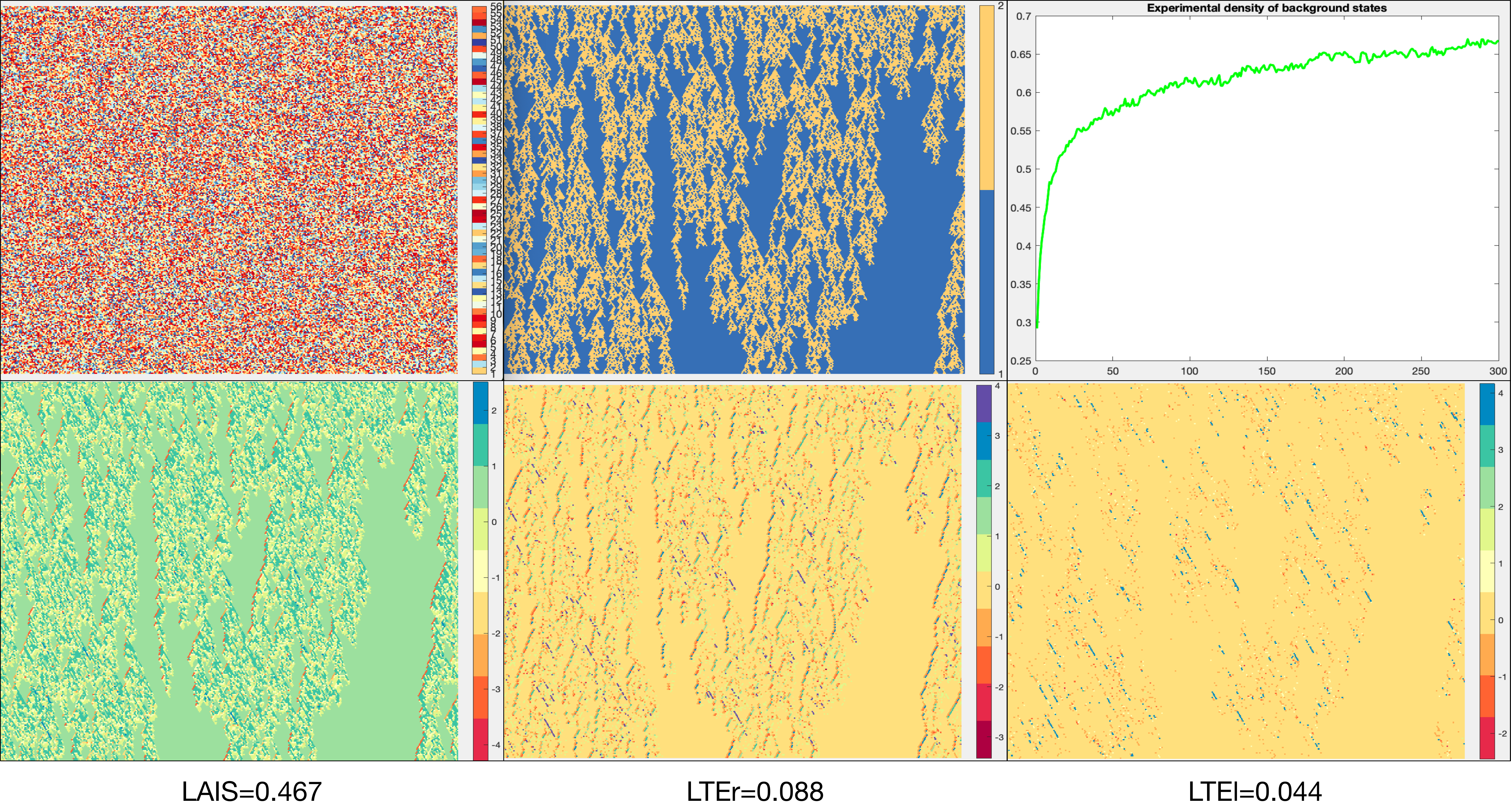}
	\caption{Randomly extended cellular automaton of $56$ states from a cellular automaton of $16$ states}
	\label{fig:AutomataComplejo_56Estados}
\end{figure}

\newpage

As more states are employed, it is more complicated to perceive the formation of gliders in a periodic background defined by multiple states. In the automaton of Figure \ref{fig:AutomataComplejo_56Estados}, we have a background specified for $ 16 $ states and extended with neighborhoods whose evolution was specified uniformly and randomly to $ 56 $ states. Notably, the use of the filter makes the appearance of the gliders that move in the periodic background clear. The experimental density is close to $ 0.65$, and the local information and transfer entropies values are again close to those observed in the emulation of Rule 110.

\begin{figure}[th]
	\centering
	\includegraphics[width=0.7\linewidth]{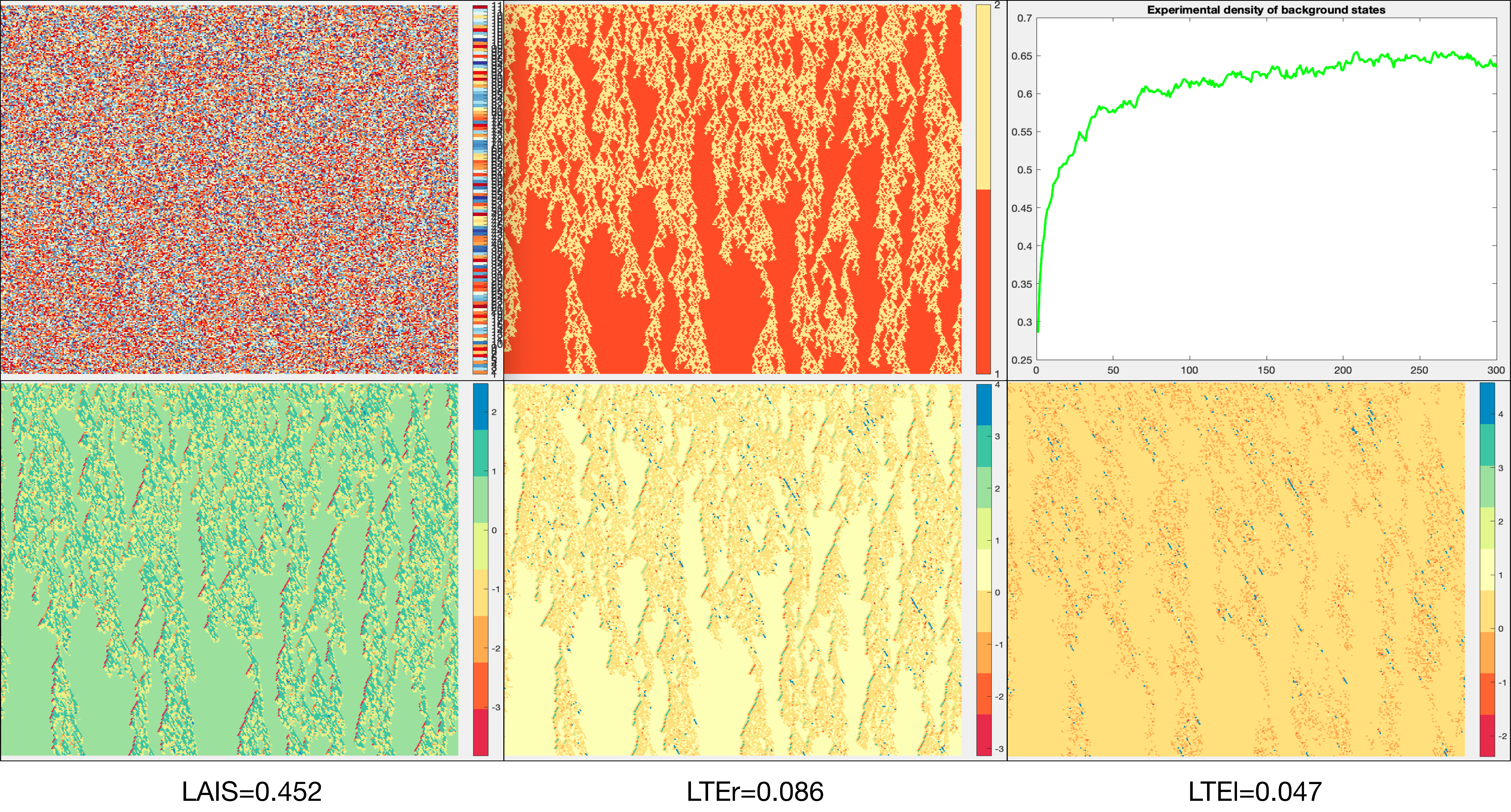}
	\caption{Randomly extended cellular automaton of $112$ states from a cellular automaton of $32$ states.}
	\label{fig:AutomataComplejo_112Estados}
\end{figure}

In Figure \ref{fig:AutomataComplejo_112Estados}, we can notice that the emergence of complex patterns can be very complicated just by inspecting the evolutions with multiple states. Moreover, we can observe this phenomenon for the automaton specified from the one which formerly had $ 32 $ states (that were used as the background) and are to be later extended to $ 112 $ states. The rise of gliders is evident in a periodic background when filtering the evolution. The experimental density of the background states is close to $ 0.6$, and the local entropies are close to those observed by the $4$-state automaton that emulates Rule 110.

The following examples demonstrate various complex cellular automata obtained through the process described above (with the change from $ 100$ to $ 400$ background states as the basis) to extended cellular automata from $ 340$ to $ 1340$ states. The evolutions display $ 500 $ cells and $ 1000 $ evolutions from first a random condition and then from a single state that is different from others. Both the options are filtered in the  background states of one color and the additional states of another in order to appreciate the appearance of the gliders in a periodic background.

\newpage

 \begin{figure}[th]
	\centering
	\includegraphics[width=0.98\linewidth]{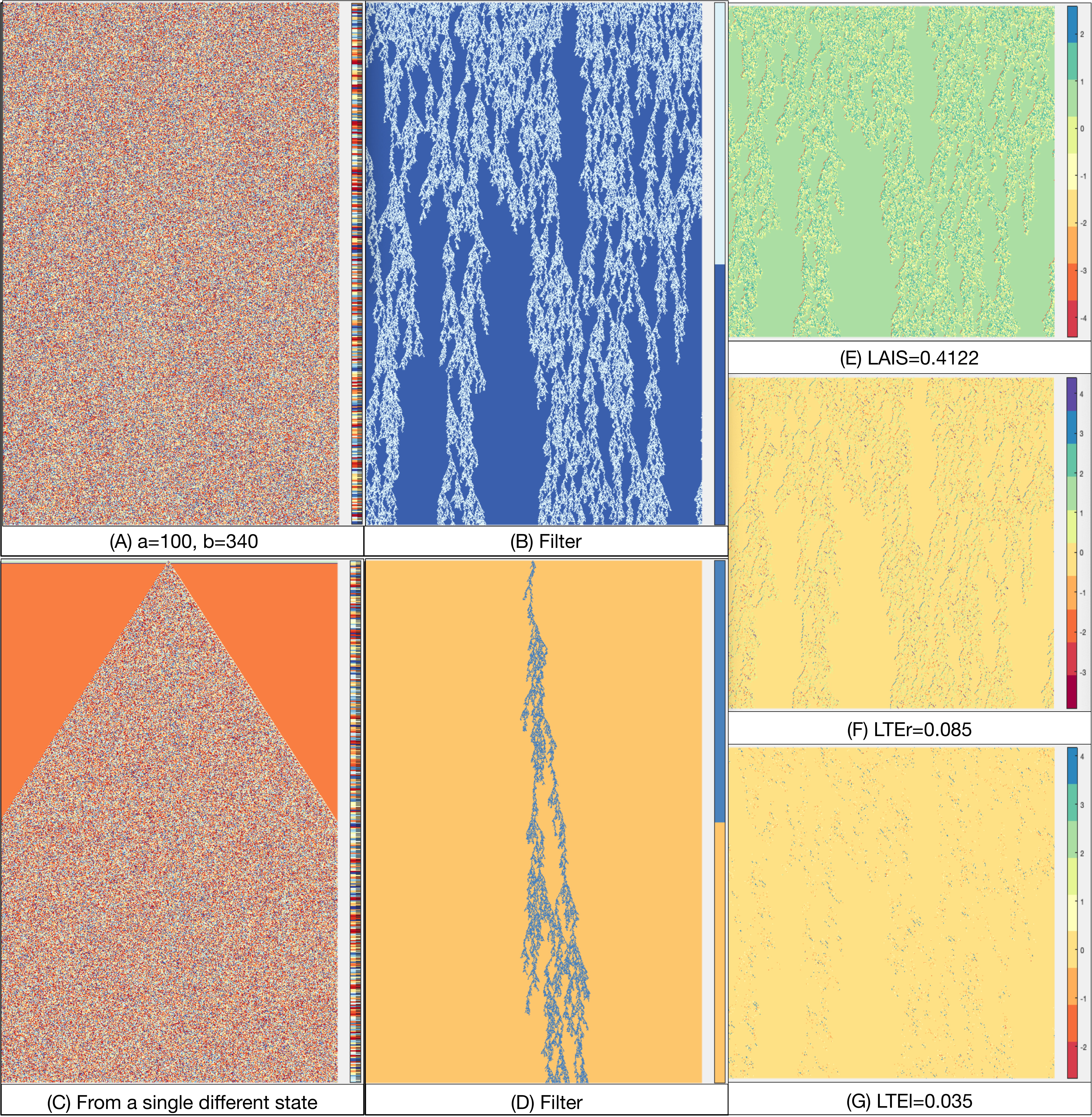}
	\caption{Cellular automta of $340$ states extended at random from a background automata of $100$ states.}
	\label{fig:AutomataComplejo_100_340Estados}
\end{figure}

\newpage

\begin{figure}[th]
	\centering
	\includegraphics[width=0.98\linewidth]{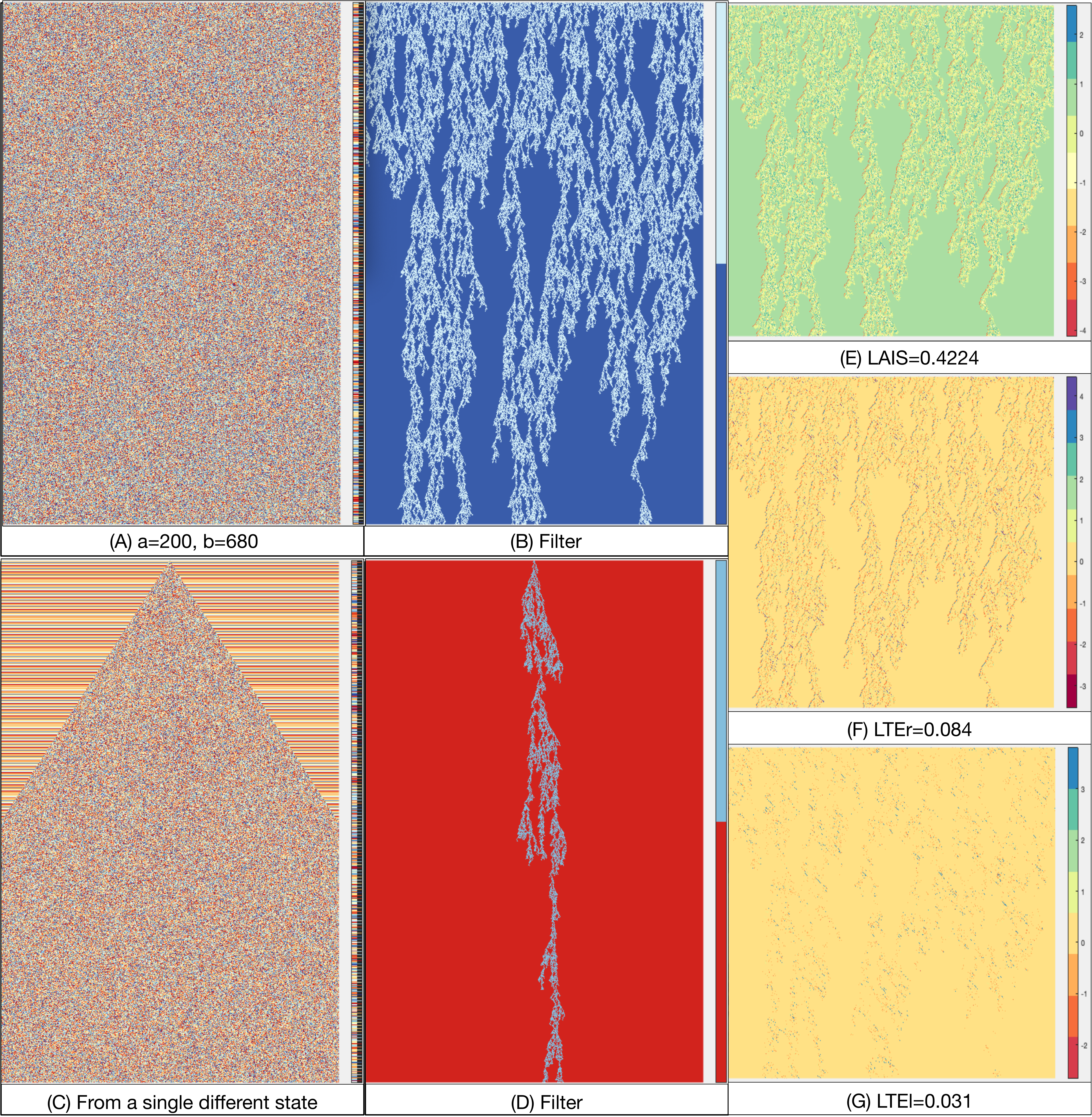}
	\caption{Cellular automta of $680$ states extended at random from a background automata of $200$ states.}
	\label{fig:AutomataComplejo_200_680Estados}
\end{figure}

\newpage

\begin{figure}[th]
	\centering
	\includegraphics[width=0.98\linewidth]{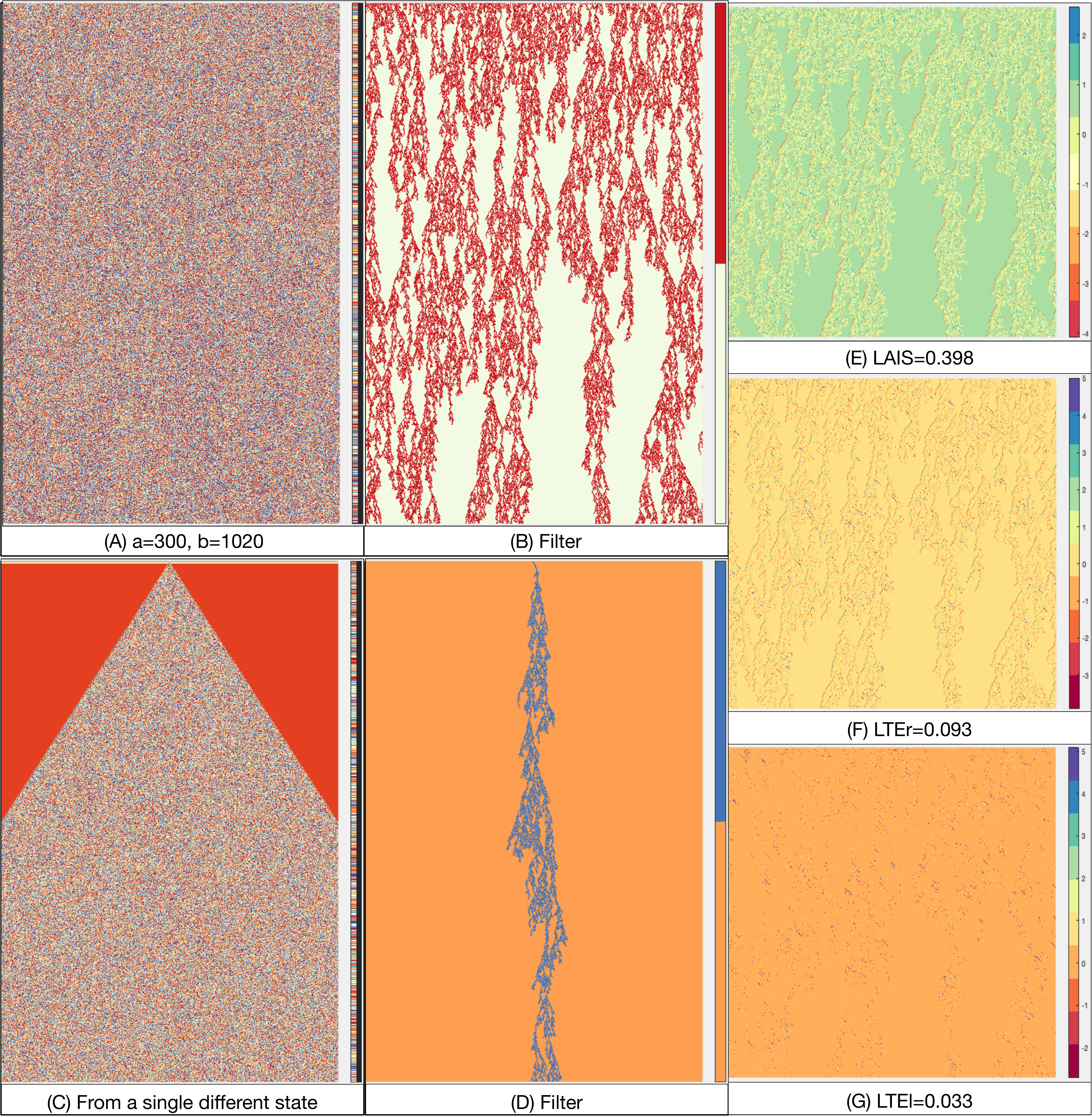}
	\caption{Cellular automta of $1020$ states extended at random from a background automata of $300$ states.}
	\label{fig:AutomataComplejo_300_1020Estados}
\end{figure}

\newpage

\begin{figure}[th]
	\centering
	\includegraphics[width=0.98\linewidth]{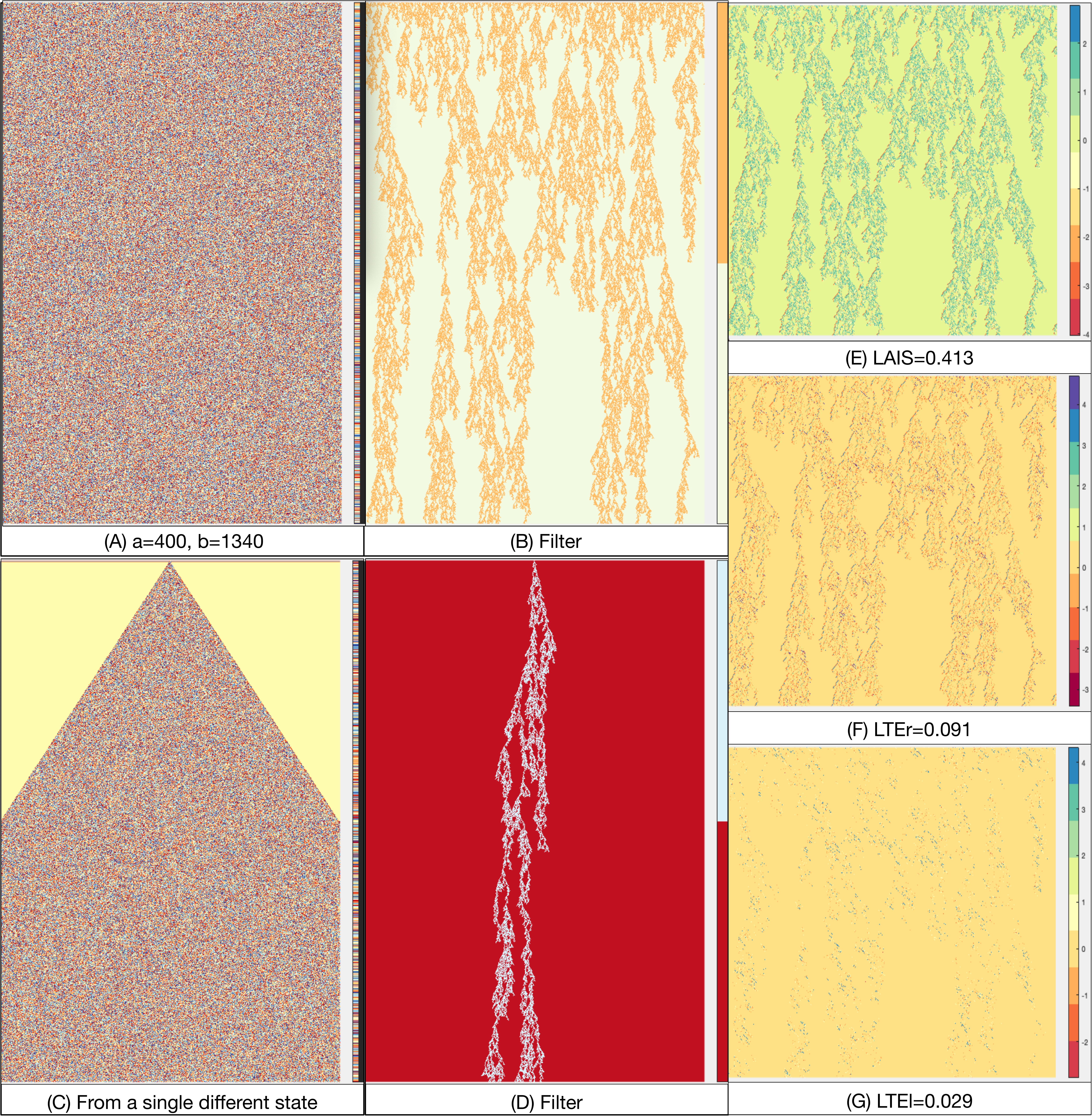}
	\caption{Cellular automta of $1340$ states extended at random from a background automata of $400$ states.}
	\label{fig:AutomataComplejo_400_1340Estados}
\end{figure}

\newpage

Table \ref{resumen_automatas_complejos} provides the summary of the characteristics of the proportion of additional states, the density of background states and the local entropies of the complex cellular automata presented in this study (Sections 4 to 6).

\begin{table}[th]
\begin{tabular}{|c|c|c|c|c|c|c|}
	\hline 
	background states & total states & $\alpha$ & density & $LAIS$ & $LTE_r$ & $LTE_l$ \\
	\hline 
	4 & 4 & 1 & 0.61 & 0.6068 & 0.087 & 0.056 \\ 
	\hline 
	4 & 12 & 2 & 0.75 & 0.459 & 0.131 & 0.058 \\ 
	\hline 
	4 & 14 & 2.5 & 0.65 & 0.000 & 0.000 & 0.000 \\ 
	\hline 
	2 & 7 & 2.5 & 0.69 & 0.574 & 0.062 & 0.039 \\ 
	\hline 
	8 & 28 & 2.5 & 0.62 & 0.479 & 0.094 & 0.035 \\ 
	\hline 
	16 & 56 & 2.5 & 0.65 & 0.467 & 0.088 & 0.044 \\ 
	\hline 
	32 & 112 & 2.5 & 0.61 & 0.452 & 0.086 & 0.047 \\ 
	\hline 
	100 & 340 & 2.4 & 0.60 & 0.412 & 0.085 & 0.035 \\ 
	\hline 
	200 & 680 & 2.4 & 0.62 & 0.422 & 0.084 & 0.031 \\ 
	\hline 
	300 & 1020 & 2.4 & 0.60 & 0.398 & 0.093 & 0.033 \\ 
	\hline 
	400 & 1340 & 2.35 & 0.63 & 0.413 & 0.091 & 0.029 \\ 
	\hline 
\end{tabular} 
\caption{\label{resumen_automatas_complejos}Density and local entropies of the previously described automata}
\end{table}

\section{Conclusions}


This study demonstrates how complexity in cellular automata can be randomly obtained by using a smaller automaton as the background and extending it to generate another larger automaton with multiple states in which gliders are formed, which interact in a periodic background. The method proposed in this study offers the possibility of generating complex cellular automata with hundreds of states.

While generating complex automata, the approximation of the density of the background states through mean-field polynomials as well as local information and transfer entropy measures are useful for specifying a simple genetic algorithm to look for evolution rules with multiple states that produce complex behaviors.

Many aspects have been identified for future researchers to explore. For instance,they can use the complexity classifications that have been presented in other works to characterize the complexity obtained through the proposed method.

In this method, only one subautomaton is specified as a generator of a periodic background. Moreover, the investigation of the application of two or more subautomata for this task has also been proposed in addition to using operations such as permutations and reflections of states on these subautomata in order to investigate whether it is possible to generate different types of dynamics and if there is a variation in the proportion of additional states needed to obtain complex behaviors.

The extension of random automata proposed in this work uses a uniform random distribution. Another task for future researchers would be to investigate other types of probabilistic distributions in order to investigate the type of complexity that can be achieved through them.
 
However, the random extension of an automaton is not the only way to obtain complex behaviors; other methods can also serve the same purpose. For example, the composition of several evolution rules may also lead to the generation of multiple state automata. However, it is relevant to know in which cases this operation may be able to produce complex behaviors.

Different measures and tools can also be applied to the ones already proposed in this study to detect and measure the complexity and produce more elaborate constructions such as certain types of gliders or glider guns, which are essential for the implementation of structures that are capable of performing computing tasks.

\section*{Acknowledgment}

This study was supported by the National Council for Science and Technology (CONACYT) with the project numbers CB-2014-237323 and CB-2017-2018-A1-S-43008, along with IPN Collaboration Network ``\,Grupo de Sistemas Complejos del IPN\,''.

\section*{References}
\bibliographystyle{elsarticle-num} 
\bibliography{References}

\end{document}